\documentclass[%
 reprint,%
 amssymb, amsmath,%
aps 
]{revtex4-1}

\usepackage{graphics}
\usepackage{amsmath}
\usepackage{amsfonts}
\usepackage{latexsym}
\usepackage{amsbsy}
\usepackage{epstopdf}
\usepackage{bm}%

\newcommand{\ket}[1]{\vert #1 \rangle}

\usepackage{color}

\usepackage{ulem} 

\newcommand{\dyadic}[1]{{#1}
\setbox0=\hbox{$\scriptstyle\leftrightarrow$}
   \setbox2=\hbox{$#1$}
   \dimen0=.5\wd0 \advance\dimen0 by-.5\wd2
   \advance\dimen0 by-\wd0
   \kern\dimen0
{^{\hbox{$\scriptstyle\leftrightarrow$}}}}

\begin{document}

\title{Atomic Spectra in a Six-Level Scheme for Electromagnetically Induced Transparency and Autler-Townes Splitting in Rydberg Atoms}
\thanks{Publication of the U.S. government, not subject to U.S. copyright.}
\author{Amy K. Robinson}
\affiliation{Department of Electrical Engineering, University of Colorado, Boulder,~CO~80305, USA}
\author{Alexandra Artusio-Glimpse}
\author{Matthew T. Simons}
\author{Christopher L. Holloway}
\email{christopher.holloway@nist.gov}
\affiliation{National Institute of Standards and Technology, Boulder,~CO~80305, USA}

\date{\today}

\begin{abstract}
We investigate electromagnetically induced transparency (EIT) and Autler-Townes (AT) splitting in Rydberg rubidium atoms for a six-level excitation scheme.  In this six-level system, one radio-frequency field simultaneously couples to two high-laying Rydberg states and results in interesting atomic spectra observed in the EIT lines.  We present experimental results for several excitation parameters.  We also present two theoretical models for this atomic system, where these two models capture different aspects of the observed spectra. One is a six-level model used to predict dominant spectral features and the other a more complex eight-level model used to predict the full characteristics of this system. Both models shows very good agreement with the experimental data.

\end{abstract}

\maketitle

\section{Introduction}

In recent years, Rydberg atoms (atoms with one or more electrons excited to a very high principal quantum number $n$ \cite{gal}) in conjunction with electromagnetically-induced transparency (EIT) techniques have been used to successfully detect and fully characterize radio frequency (RF) electric (E) fields \cite{emcconf}-\cite{access}.   The E-fields are detected using EIT both on resonance as Autler-Townes (AT) splitting and off-resonance as ac Stark shifts. This approach allows for the detection of the  amplitude, phase, and polarization of E-fields and modulation signals, all in one compact atomic-vapor cell.

This EIT field sensing (and Rydberg atom generation) approach is typically based on four-level ladder excitation schemes,  Fig.~\ref{EIT}(a), which include a ground-state probe laser (levels $\ket{1}$ and $\ket{2}$), a coupling laser generating Rydberg states (levels $\ket{2}$ and $\ket{3}$), and an RF source to couple two Rydberg states (levels $\ket{3}$ and $\ket{4}$). The addition of the resonant radio frequency (RF) field (labeled as RF1) results in AT splitting. The AT split gives a direct SI-traceable measurement of the RF E-field strength \cite{emcconf, r4, r3, r2}. The SI-traceability stems from the fact that, in this approach, the E-field strength is expressed in terms of Planck's constant, where Planck's constant is now an SI defined constant in the redefinition of the SI that went into effect last year \cite{si1, si2}.  When the RF field is off resonant of a Rydberg transition, a three-level scheme is used (the same first three states used in Fig. \ref{EIT}{(a), i.e., $\ket{1}$, $\ket{2}$, and $\ket{3}$)  and the E-field is detected by monitoring the ac Stark shift of this three-level EIT line.

\begin{figure}
\centering
\scalebox{.3}{\includegraphics*{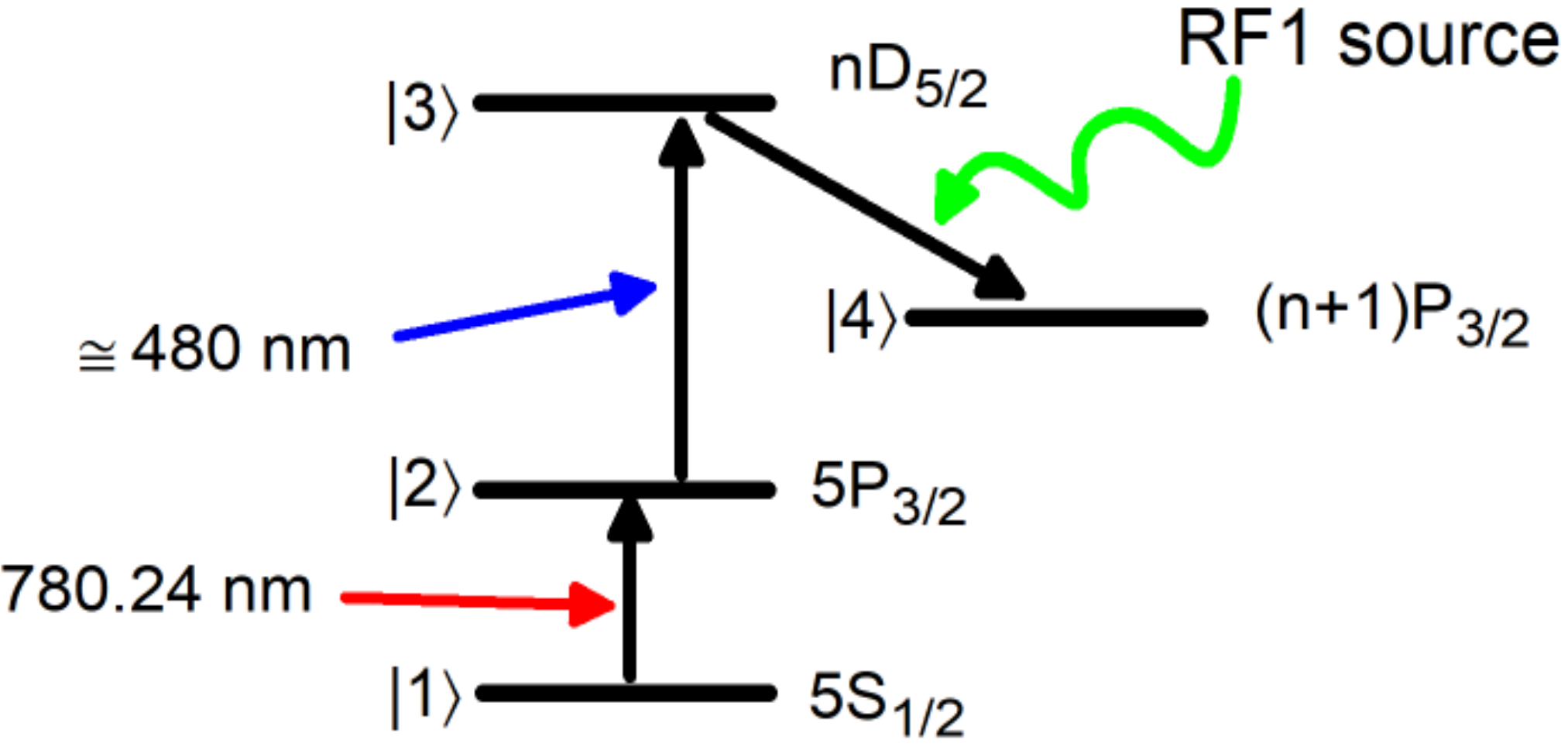}}\\
\begin{centering} (a) four-level system \end{centering}\\ \vspace{2mm}
\scalebox{.4}{\includegraphics*{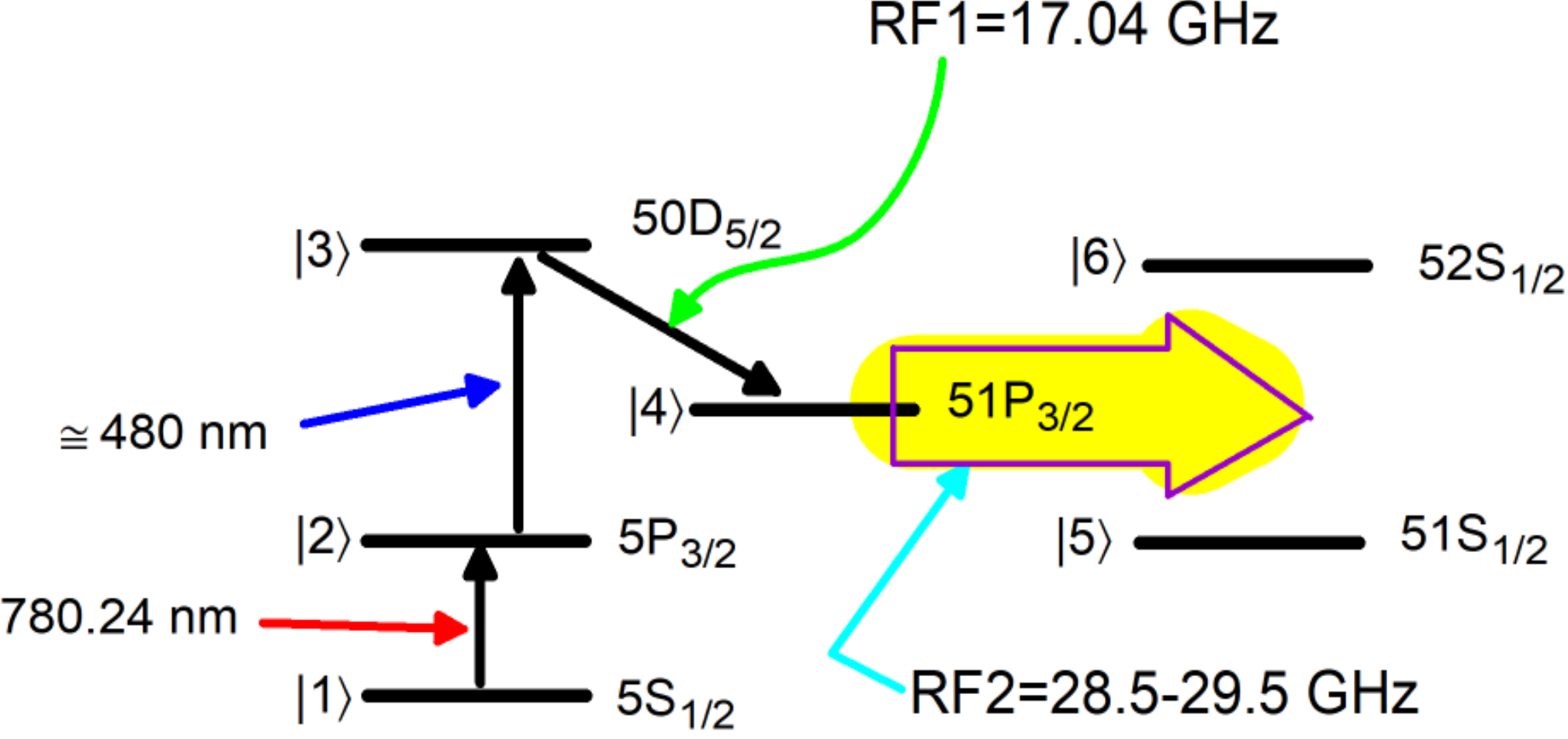}}\\
\begin{centering} (b) six-level system \end{centering}\\
\caption{EIT schemes: (a) four-level ladder scheme, and (b) six-level scheme. For the six-level scheme, the different in the transition frequencies between $\ket{4}$-$\ket{5}$ (28.92~GHz) and $\ket{4}$-$\ket{6}$ (29.24~GHz) is 324.8~MHz.}
\label{EIT}
\end{figure}

The research over the past ten years into this four-level scheme has allowed us to control ensembles of room-temperature atoms in such a manner that we are able to develop interesting and unique applications of these Rydberg atom-based sensors. Besides SI traceable E-field probes, other applications range from atom-based receivers to imaging capabilities, among many others. Because of the numerous potential applications of this new sensor technology, several groups around the world have begun programs in the area of Rydberg atom-based detectors/sensors/receivers (including universities, private companies, government agencies and most of the national metrology institutes around the world).

The success of the four-level EIT scheme entices one to look for other EIT schemes that might allow for additional sensing capabilities, as well as to evaluate atom-based receivers in the presence of multiple RF signals. In this paper, we explore the atomic spectra of the six-level EIT scheme shown in Fig.~\ref{EIT}(b). The first four levels of our six-level scheme use the same ladder schemes described above, while the fifth and six levels ($\ket{5}$ and $\ket{6}$) correspond to two additional Rydberg states [coupled to by one additional RF field (RF2)]. The six-level EIT scheme presents interesting atomic spectra that are not accessible with the basic four-level system. For example, the transition frequency for $\ket{4}$-$\ket{5}$ is 28.92~GHz and the transition frequency for $\ket{4}$-$\ket{6}$ is 29.24~GHz, a difference of only 324.8~MHz. Because the two high laying Rydberg states ($\ket{5}$ and $\ket{6}$) are relatively close in energy levels to the Rydberg state $\ket{4}$, one RF field (the RF2 field) can couple to two states simultaneously ($\ket{4}$-$\ket{5}$ and $\ket{4}$-$\ket{6}$). This simultaneously coupling of the two high laying Rydberg states results in the interesting, complex atomic spectra shown in experimental and modeling results sections of this paper.   Furthermore, we will see that since we chose P-to-S transitions to couple to the $\ket{5}$ and $\ket{6}$ states, some magnetic pathways do not couple adding to the richness of the observed behavior.

The paper shows experimental results for EIT spectra for this six-level system for various parameters. We also present two models to predict the observed behavior. In fact, we will see that to fully represent all the features in our experimental data, a multiple-level model (including magnetic sublevels and the fine- and hyperfine-structures) is required. Here, we present a six-level and eight-level model. The six-level model captures the main features observed when coupling to two states simultaneously. The eight-level model includes these features, but also includes the magnetic sublevel behavior. These models and experimental results show very good agreement to one another.  The results of this investigation can open up new schemes for field sensing with Rydberg atoms, and we discuss possibilities.

\section{Experimental Setup}

The experimental setup is depicted in Fig.~\ref{setup}, which consists of a 780~nm probe laser, a 480~nm coupling laser, two RF signal generators (SG), two horn antennas, a photodetector connected to an oscilloscope, and a 25 mm diameter cylindrical glass vapor cell of 75 mm length filled with rubidium ($^{85}$Rb) atom vapor. The atomic states used in these six-level EIT experiments are shown in Fig.~\ref{EIT}(b) and  correspond to the $^{85}$Rb  $5S_{1/2}$ ground state,  $5P_{3/2}$ excited state, and four Rydberg states ($\ket{3}\Rightarrow 50D_{5/2}$, $\ket{4}\Rightarrow 51P_{3/2}$, $\ket{5}\Rightarrow 51S_{1/2}$, and $\ket{6}\Rightarrow 52S_{1/2}$).

\begin{figure}
\centering
\scalebox{.09}{\includegraphics*{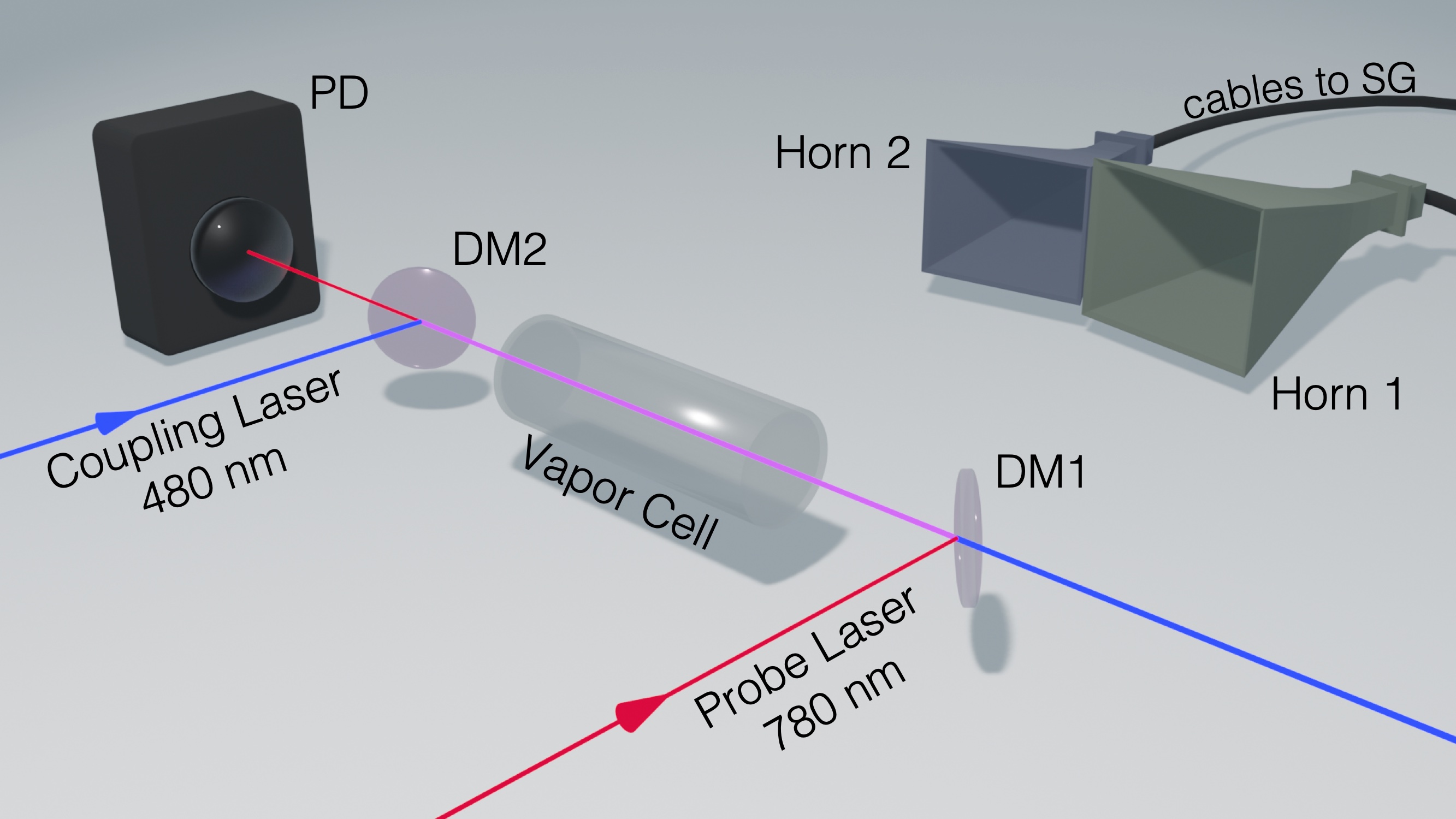}}\\
\caption{Experimental setup for six-level EIT scheme. In this figure, PD refers to the photodetector and DM1 and DM2 refers to dichroic mirrors.}
\label{setup}
\end{figure}

The probe laser is locked to the D2 transition (\mbox{$5S_{1/2}(F=3)$~--~$5P_{3/2}(F=4)$} or wavelength of \mbox{$\lambda_p=780.24$~nm} \cite{stackrb}).  To produce an EIT signal, we apply a counter-propagating coupling laser with $\lambda_c \approx 480.1$~nm and scan it across the $5P_{3/2}$-$50D_{5/2}$ Rydberg transition. We use a lock-in amplifier to enhance the EIT signal-to-noise ratio by modulating the coupling laser amplitude with a 37~kHz square wave. This removes the background and isolates the EIT signal.  RF sources are used to couple to the Rydberg states $\ket{4}$, $\ket{5}$, and $\ket{6}$. To generate these RF fields, the output of two signal generators (SG) are connected to two different horn antennas (referred to as Horn~1 and Horn~2).
Each antenna is placed 40~cm from the laser beam locations in the vapor cell. One SG and Horn~1 are used at 17.04~GHz to couple Rydberg states $50D_{5/2}$ and $51P_{3/2}$. The other SG and Horn 2 are used to couple to two different Rydberg states ($51P_{3/2}$-$51S_{1/2}$ with a transition frequency of 28.92~GHz  and $51P_{3/2}$-$52S_{1/2}$ with a transition frequency of 29.24~GHz), where the frequency of the SG ranges from 28.5~GHz to 29.5~GHz during the experiments. This allows us to couple the two states ($51P_{3/2}$-$51S_{1/2}$ and $51P_{3/2}$-$52S_{1/2}$) separately or simultaneously, depending on the frequency used. Note that in our case, the difference in the on-resonance transition frequencies for  $\ket{4}$-$\ket{5}$ and $\ket{4}$-$\ket{6}$ are only 324.8~MHz.  The RF power levels for both the SGs were varied in these experiments and the various values are stated below. The antenna gains for Horn~1 and Horn~2 relative to an isotropic antenna are 50.1 (or $17$~dBi) and 79.4 (or $19$~dBi), respectively.  In these experiments, the optical beams and the RF electric fields are co-linearly polarized.

The Rabi frequency between the various states is defined as (in units of rad/s):
\begin{equation}
\Omega_{p,c,RF1,RF2}=\frac{d \, e\,a_o}{\hbar}\,|E|\,\,\, ,
\label{e1}
\end{equation}
where $\hbar$ is Planck's constant, $d$ is the normalized atomic dipole moment (normalized by $e\,a_o$ and includes both the angular and radial parts), $e$ is the elementary charge (in units of C), $a_0$ is the Bohr radius (in units of m), and $|E|$ is the magnitude of the electric field (in units of V/m) for the various sources.
For the optical fields, $|E|$ is given by
\begin{equation}
|E|=\sqrt{\frac{8\,{\rm ln}2}{\pi\,c\,\epsilon_o}}\frac{\sqrt{P}}{D_{FWHM}}\,\,\, ,
\label{e2}
\end{equation}
where $P$ is the laser power (in units of W), $D_{FWHM}$ is the full-width at half maximum (FWHM) of the laser beams, $\epsilon_o$ in the permittivity of free-space, and $c$ is the speed of light {\it in vacuo}.
For the RF fields, $|E|$ is given by
\begin{equation}
|E|={\cal F}\frac{1}{\sqrt{2\,\pi\,c\,\epsilon_o}}\frac{\sqrt{P\, G}}{R} \,\,\, ,
\label{e3}
\end{equation}
where $P$ is the input power to the horn antenna (in units of W), $G$ is the on-axis gain of the horn antenna, and $R$ is the distance from the horn antenna to the laser beams in the vapor cell (in units of m). Finally, ${\cal F}$ is the cell perturbation factor. Because the vapor cell is a dielectric, the RF fields can exhibit multi-reflections inside the cell and RF standing waves (or resonances) in the field strength can develop inside the cell \cite{r2, r5, emcpaper, fan2}. Thus, for a given location inside the cell, the RF field inside the cell can be larger or smaller than incident field. The parameter ${\cal F}$ accounts for this effect, and it can be determined numerically or experimentally \cite{r2, r5, emcpaper, fan2}.  The RF field distribution at $17.04$~GHz for the cell used in these experiments are shown in Fig.~9 of \cite{r5}. Using these results we estimate ${\cal F}\approx0.5$ for 17.04~GHz.  Using a same numerical code as that discussion in \cite{r5}, we estimate ${\cal F}\approx0.5$ for 28.5~GHz.

The normalized atomic-dipole moment ($d$) for the various transitions are as follows. For the $5S_{1/2}$-$5P_{3/2}$ transition, $d=1.93$ (obtained from Table~7 and Table~10 in \cite{stackrb}, where for the angular part, we averaged over all seven of the $m_F$ magnetic $\pi$ transitions). For the $5P_{3/2}$-$50D_{5/2}$ transition, $d=0.0099$ (obtained from a numerically calculated radial part of 0.0222 and an angular part of 0.445 by averaging over the four $m_J$ magnetic $\pi$ transitions). Note that $m_J$ is the ``good'' magnetic quantum number for the Rydberg states. For the $50D_{5/2}$-$51P_{3/2}$ transition, $d=1430.4$ (obtained from a numerically calculated radial part of 3214.33 and an angular part of 0.445 by averaging over the four $m_J$ magnetic $\pi$ transitions).  For the $51P_{3/2}$-$51S_{1/2}$ transition, $d=1282.4$ (obtained from a numerically calculated radial part of 2617.67 and an angular part of 0.48989, the $m_J=\pm 1/2$ magnetic $\pi$ transitions). For the $51P_{3/2}$-$52S_{1/2}$ transition, $d=1250.77$ (obtained from a numerically calculated radial part of 2553.13 and an angular part of 0.48989, the $m_J=\pm 1/2$ magnetic $\pi$ transitions).
Note, coupling to states $\ket{5}$ and $\ket{6}$ are P-to-S transitions, therefore only the $m_J=\pm 1/2$ magnetic transitions are allowed for the $\pi$ transition. As we will see below, this gives some magnetic pathways that do not couple to the P-to-S Rydberg transitions. Also note that the transitions $\ket{4}$-to-$\ket{5}$ and $\ket{4}$-to-$\ket{6}$ have very similar values for $d$ and the difference between the transition frequencies is 324.8~MHz apart. This allows for simultaneously coupling to the two Rydberg states with one RF source and results in interesting EIT spectra as the frequency of RF2 is varied (or detuned).


\section{Experimental Results}

Fig.~\ref{eitsignal} shows the EIT signal with no RF1 and no RF2 fields as a function of the coupling laser detuning ($\Delta_c$), i.e., using only the probe and coupling lasers. The results where obtained for a probe laser focused to a full-width at half maximum (FWHM) of 80~$\mu$m, with a power of 374~nW and a coupling laser with power of 82~mW, focused to a FWHM of 110~$\mu$m. Using eqs.~(\ref{e1}) and (\ref{e2}), along with the values of $d$ given above, the Rabi frequencies for the probe and coupling lasers are $\Omega_p=2\pi\times$(4.8~MHz) and $\Omega_c=2\pi\times(8.5$~MHz), respectively.

\begin{figure}
\vspace{5mm}
\centering
\scalebox{.3}{\includegraphics*{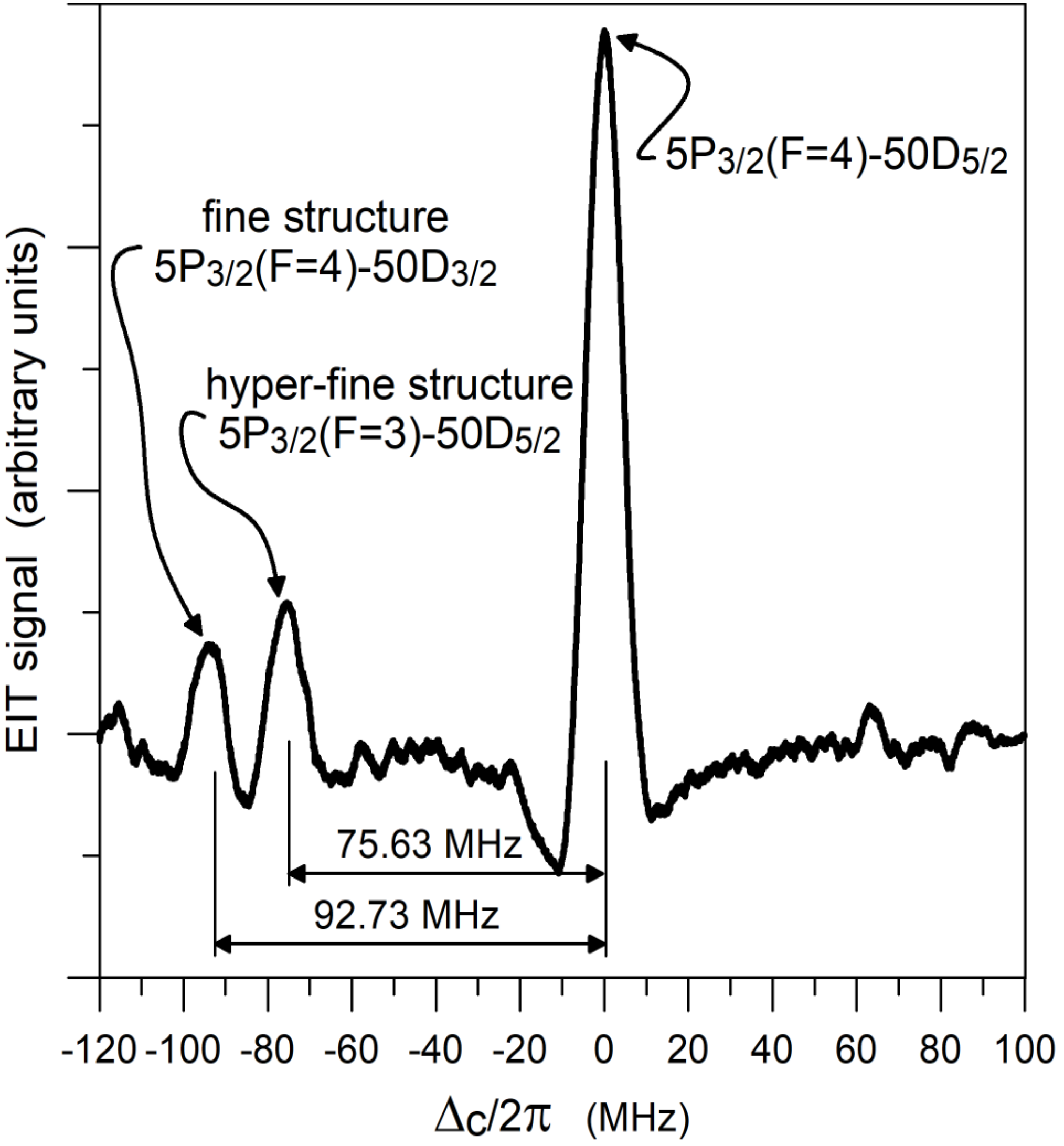}}\\
\caption{Experimental EIT signal for the first three levels ($\ket{1}$, $\ket{2}$ and $\ket{3}$ shown in Fig.~\ref{EIT}(b), i.e., no RF1 and no RF2 fields.}
\label{eitsignal}
\end{figure}

The main-central peak at \mbox{$\Delta_c=0$} corresponds to the $5P_{3/2}$(F=4)-$50D_{5/2}$. The peak at \mbox{$\Delta_c/2\pi=-75.63$~MHz} corresponds the hyperfine structure transition $5P_{3/2}$(F=3)-$50D_{5/2}$. Since the coupling laser is scanned, the separation of 75.63~MHz is determined by \mbox{$120.96(\frac{\lambda_p}{\lambda_c}-1)$~MHz}, where $\lambda_p$ and $\lambda_c$ are the wavelengths of the probe and coupling lasers, and 120.96~MHz is the hyper-fine structure separation between $F=3$ and $F=4$ \cite{stackrb}. The peak at $\Delta_c/2\pi=-92.73$~MHz corresponds the the fine-structure transition $5P_{3/2}$(F=4)-$50D_{3/2}$.

\begin{figure}
\centering
\scalebox{.26}{\includegraphics{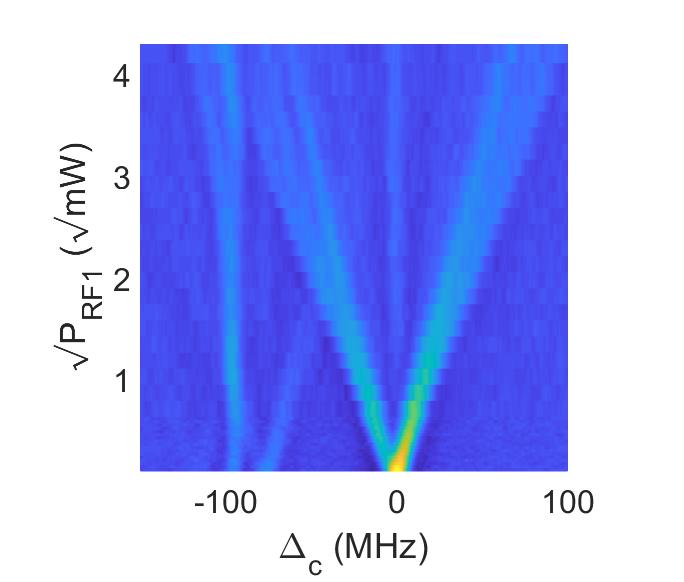}}\\
\vspace*{-2mm}
{\hspace{-1mm}\tiny{(a)}}\\
\vspace*{1mm}
\scalebox{.26}{\includegraphics{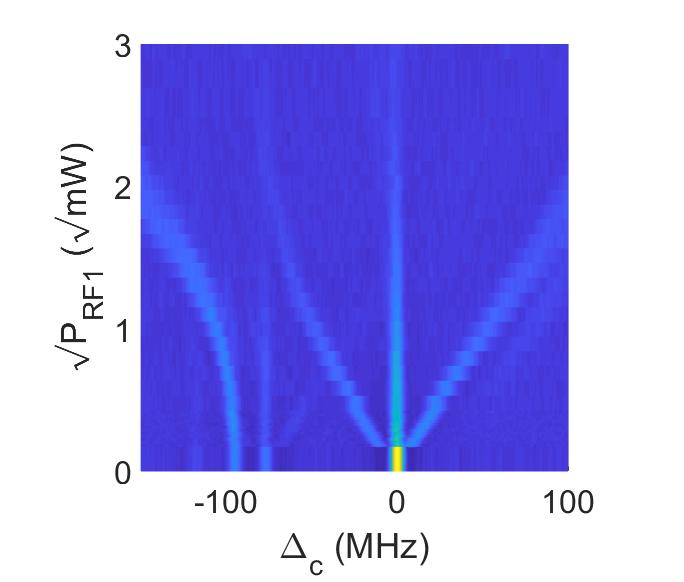}}\\
\vspace*{-2mm}
{\hspace{-1mm}\tiny{(b)}}\\
\vspace{1mm}
\scalebox{.26}{\includegraphics{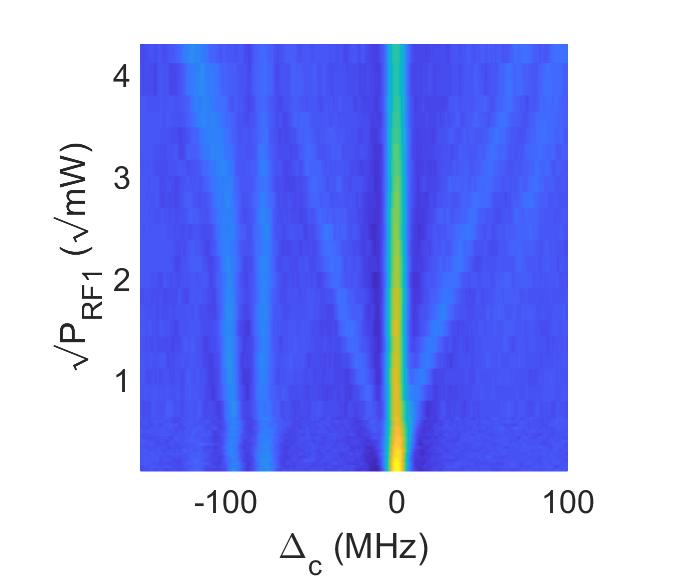}}\\
\vspace*{-2mm}
{\hspace{-1mm}\tiny{(c)}}\\
\caption{Experimental EIT signal for the six-level systems shown in Fig.~\ref{EIT}(b); (a) $P_{RF2}$ off, (b) $P_{RF2}=3.16$~mW [$\Omega_{RF2}=2\pi\times$(78~MHz)], and (c)  $P_{RF2}=10$~mW [$\Omega_{RF2}=2\pi\times$(138~MHz)].}
\label{rf1ps}
\end{figure}

Fig.~\ref{rf1ps} shows the results for scanning the power of the RF1 source for different RF2 power levels. Throughout the paper, the stated power levels for RF1 and RF2 refer to the input power level to the horn antennas. For these results the Rabi frequency for the probe and coupling laser are both $\Omega_{p,c}=2\pi\times$(2~MHz) [probe laser power of $1.6~\mu$W focused to $400~\mu$m and a coupling laser power of 92~mW, focused to a FWHM of 480~$\mu$m].   In these figures, the x-axis is the detuning of the coupling laser ($\Delta_c$), the y-axis is the power level of RF1, and the shaded contour is the relative amplitude of the EIT lines. The EIT lines starting at $\Delta_c=0$ and $P_{RF1}=0$ correspond to $5P_{3/2}$(F=4)-$50D_{5/2}$ transitions, the EIT lines starting at $\Delta_c/2\pi=-75.57$~MHz and $P_{RF1}=0$ is the hyper-fine structure transition corresponding to the \mbox{$5P_{3/2}$(F=3)-$50D_{5/2}$}, and the EIT line starting at $\Delta_c/2\pi=-92.73$~MHz and $P_{RF1}=0$ is the fine-structure transition corresponding to the $5P_{3/2}$(F=4)-$50D_{3/2}$. The main EIT line at ($5P_{3/2}$(F=4)-$50D_{5/2}$) and the hyper-fine structure line ($5P_{3/5}$(F=3)-$50D_{5/2}$) experience AT splitting when the RF1 power is increased. The hyper-fine structure splitting is clearly seen in Fig.~\ref{rf1ps}(a). For low RF1 power levels, the AT splitting is linear (linear with $|E|$ or $\sqrt{P_{RF1}}$) and becomes quadratic (varies as $|E|^2$) for high RF1 power levels ($\sqrt{P_{RF1}}>0.5~\sqrt{{\rm mW}}$ or $P_{RF1}>0.3$~mW) due to the ac Stark shift.  On the other hand, the fine structure EIT line ($5P_{3/2}$(F=4)-$50D_{3/2}$) does not experience AT splitting when RF1 is applied. However, we do see that this fine-structure EIT line does experience ac Stark shifts, seen especially for $P_{RF1}>1.0$~mW.
When RF2 is applied for all RF1 power levels, we see straight,  predominant unshifted features, EIT lines along the $\Delta_c=0$ axis ($51P_{3/2}$-$52S_{1/2}$) and along $\Delta_c/2\pi=75.57$~MHz (this is the hyperfine line of the $51P_{3/2}$-$52S_{1/2}$ transition).

Fig.~\ref{rf1d2} shows the EIT spectra for detuning RF1 ($\Delta_{RF1}$) for three different RF1 power levels and no RF2 power. These results are for the same optical Rabi frequencies as those used in Fig.~\ref{rf1ps} [i.e., the Rabi frequency for the probe and coupling laser are both $\Omega_{p,c}=2\pi\times$(2~MHz)]. Once again, we see that the main EIT line experiences AT splitting and the fine-structure does not experience splitting.  From these plots we also see the splitting of the hyper-fine structure.  These results show the usual AT behavior with RF detuning, in that there are two main effects on the observed splitting of the EIT signal \cite{r16}. First, the two peaks of the EIT signal are non-symmetric (i.e., the heights of the two peaks are not the same). Second, the separation between the two AT peaks increases with RF detuning.

\begin{figure}
\centering
\vspace{-2mm}
\scalebox{.26}{\includegraphics{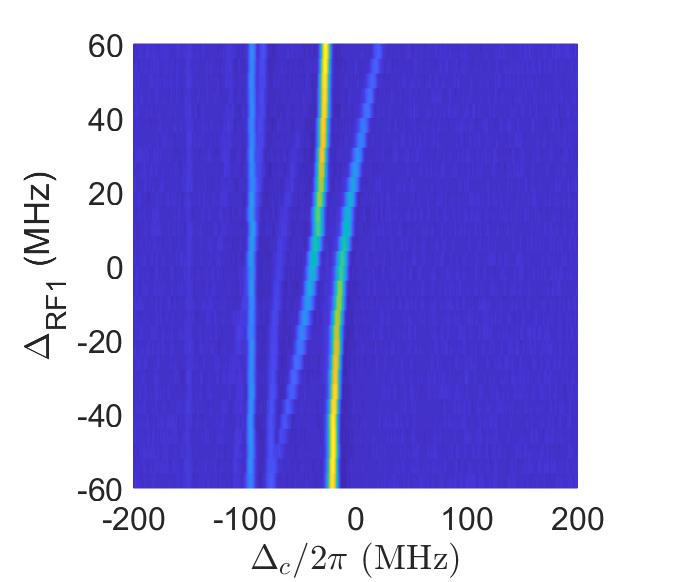}} \\
\vspace{-2mm}
{\hspace{-1mm}\tiny{(a)}}\\
\scalebox{.26}{\includegraphics{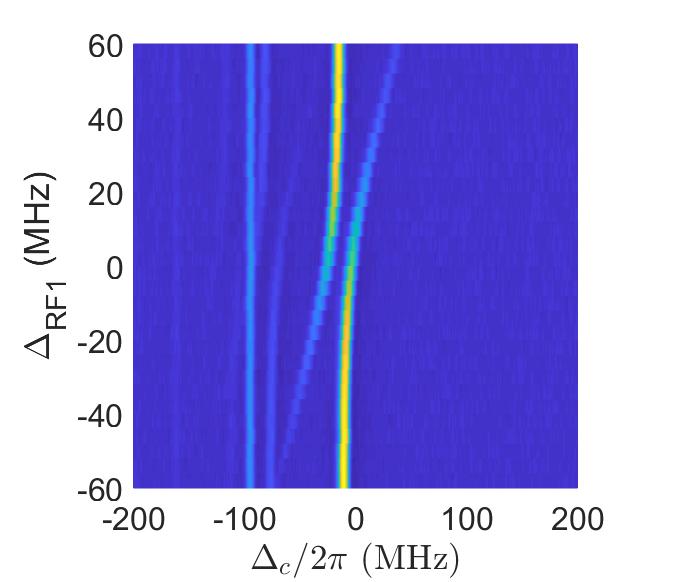}}\\
\vspace{-2mm}
{\hspace{-1mm}\tiny{(b)}}\\
\scalebox{.26}{\includegraphics{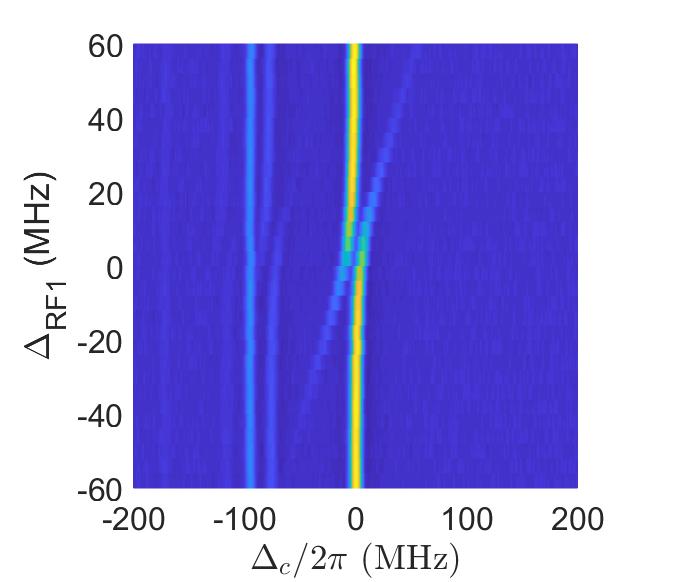}}\\
\vspace{-2mm}
{\hspace{-1mm}\tiny{(c)}}\\
\caption{Experimental EIT signal for the four-level systems shown in Fig.~\ref{EIT}(b) (i.e., no RF2 power):  (a) $P_{RF1}=0.16$~mW [$\Omega_{RF1}=2\pi\times$(32~MHz)], (b) $P_{RF1}=0.08$~mW [$\Omega_{RF1}=2\pi\times$(11~MHz)] and (b)  $P_{RF1}=0.03$~mW [$\Omega_{RF1}=2\pi\times$(6~MHz)].}
\label{rf1d2}
\end{figure}

The next set of data, Fig.~\ref{rf2power}, show the EIT spectra for scanning the RF2 power levels for four different RF1 power levels. These results are for an RF2 frequency of $f_{RF2}=$28.92~GHz, the on resonant $51P_{3/2}$-$51S_{1/2}$ transition frequency, with the same optical Rabi frequencies as those in used in Fig.~\ref{eitsignal} (i.e., the Rabi frequency for the probe and coupling laser are $\Omega_p=2\pi\times$(4.8~MHz) and $\Omega_c=2\pi\times$(8.5~MHz), respectively).
In this data set we scan RF2 power over several orders of magnitude; therefore, the x-axis is on a log-scale in which ``dBm'' is defined as ${10Log_{10}(P_n)}$, where $P_n$ is the power level normalized by 1~mW.
It is interesting to observe that at certain $P_{RF2}$ values (for example around $P_{RF2}\approx1$~mW [or 0~dBm] for $P_{RF1}=1$~mW) a strong EIT line at $\Delta_c=0$ begins to appear (this is a results of the $51P_{3/2}$-$51S_{1/2}$ transition). Around that same power level, we see additional lines that start to curve off the AT lines. For example, from Fig.~\ref{rf2power}{b} ($P_{RF1}=1$~mW), one curves to the left of the AT line at $\Delta_c/2\pi=-20$~MHz and one curves to the right of the AT line located at $\Delta_c/2\pi=20$~MHz. As the EIT line at $\Delta_c=0$ becomes stronger, the AT lines at $\Delta_c/2\pi=\pm20$~MHz become weaker. However, the AT line at $\Delta_c/2\pi=\pm20$~MHz does not completely disappear. This is a result of using P-to-S transition for $\ket{4}$-to-$\ket{5}$.  The $m_J=\pm3/2$ magnetic sublevels in state $51P_{3/2}$ are not coupled to the $51S_{1/2}$ level by the $28.92$~MHz field since we are using a $\pi$ transition.  This is more apparent in the next set of data (Fig.~\ref{rf2detuning}). This same behavior is seen in the other curves in Fig.~\ref{rf2power}, the only difference being where the AT peaks are located for the other $P_{RF1}$ power levels (i.e., $\Delta_c/2\pi=\pm11$~MHz for $P_{RF1}=0.32$~mW, $\Delta_c/2\pi=\pm35$~MHz for $P_{RF1}=3.16$~mW, and $\Delta_c/2\pi=\pm63$~MHz for $P_{RF1}=10$~mW).

The measured AT splitting (or the RF Rabi frequency for RF1) shown in Fig.~\ref{rf2power}(b) is approximately 40~MHz.  The results in Fig.~\ref{rf2power}(b) are for $P_{RF1}=1$~mW. Using this power level, the parameter ${\cal F}\approx0.5$, and the expression in (\ref{e1}) and (\ref{e3}), we calculate the Rabi frequencies to be 39.7~MHz. This is in good agreement with the measured value. The other measured AT splitting in Fig.~\ref{rf2power} are in as equal agreement with calculated values. The calculated values for the Rabi frequencies are shown in the figure caption. This is important, in that, while we used numerical calculations to determine ${\cal F}$, we could have equally well used the experimental values of the AT splitting (RF Rabi frequency) to determine ${\cal F}$. That is, a series of measured Radi frequencies can be used to calibrate the vapor cell (i.e., determine ${\cal F}$)

\begin{figure*}
\centering
\scalebox{.26}{\includegraphics{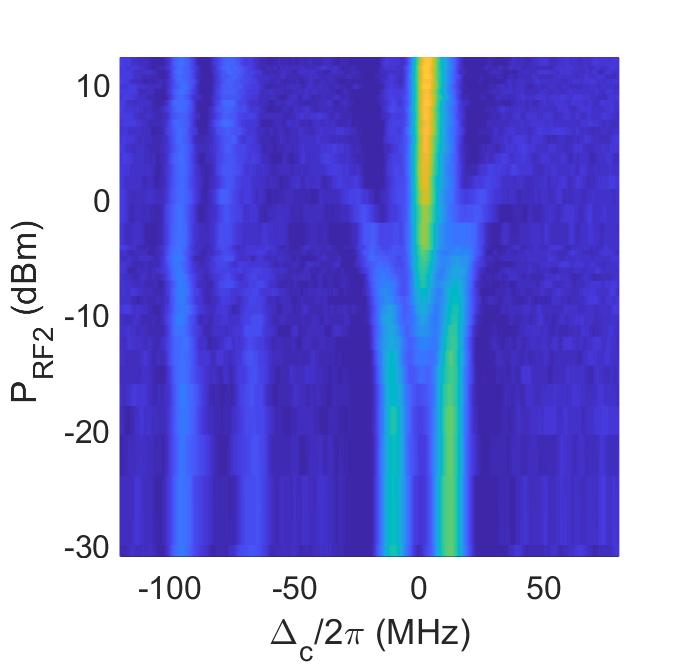}} \hspace{2mm}
\scalebox{.26}{\includegraphics{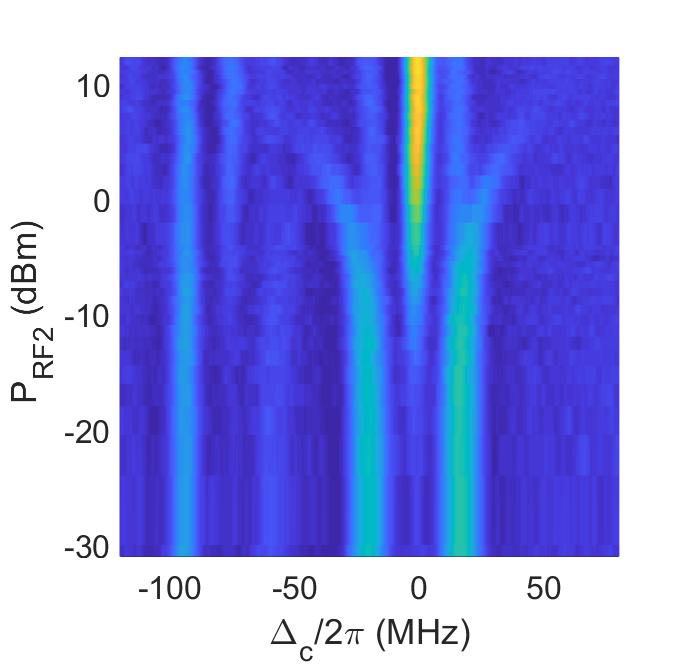}}\\ \vspace*{-2mm}
{\hspace{-1mm}\tiny{(a) \hspace{77mm} (b)}}\\
\vspace*{2mm}
\scalebox{.26}{\includegraphics{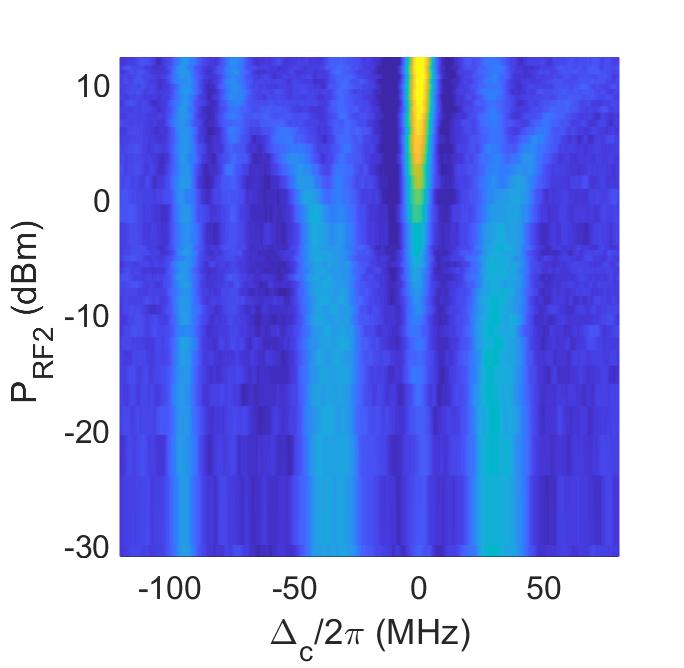}} \hspace{2mm}
\scalebox{.26}{\includegraphics{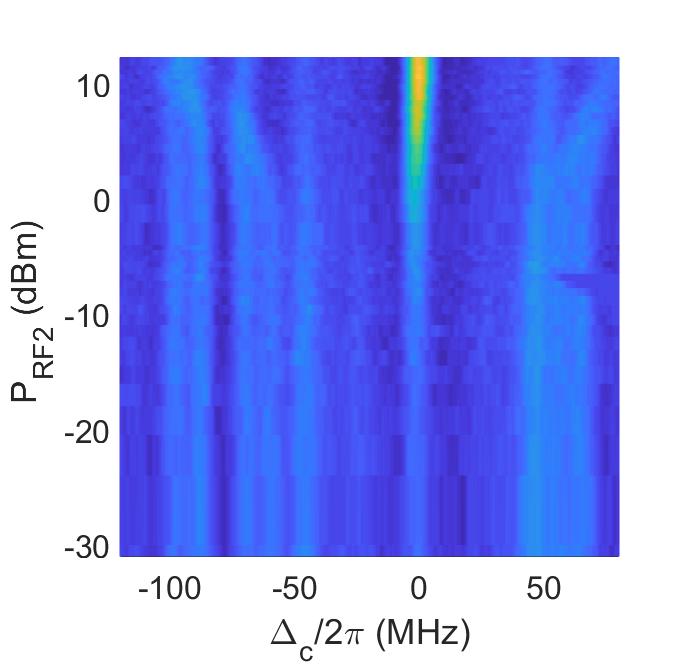}}\\ \vspace*{-2mm}
{\hspace{-1mm}\tiny{(c) \hspace{77mm} (d)}}\\
\caption{Experimental EIT signal for the six-level systems shown in Fig.~\ref{EIT}(b) for scanning the power levels of RF2; (a) $P_{RF1}=0.32$~mW [$\Omega_{RF1}=2\pi\times$(22~MHz)], (b) $P_{RF1}=1$~mW [$\Omega_{RF1}=2\pi\times$(40~MHz)], (c) $P_{RF1}=3.16$~mW [$\Omega_{RF1}=2\pi\times$(71~MHz)], and (d)  $P_{RF1}=10$~mW [$\Omega_{RF1}=2\pi\times$(126~MHz)].}
\label{rf2power}
\end{figure*}

Finally, some of the more interesting EIT spectra are obtain for detuning RF2 ($\Delta_{RF2}$) and the results are shown in Fig.~\ref{rf2detuning} (the left column are the experimental results). These results are for $P_{RF1}=1$~mW [$\Omega_{RF1}=2\pi\times$(40~MHz)] and for the Rabi frequencies for the probe and coupling laser of $\Omega_p=2\pi\times$(4.8~MHz) and $\Omega_c=2\pi\times$(8.5~MHz), respectively. In these plots, $\Delta_{RF2}=0$ is relative to $29.08$~GHz. This is the average of on-resonant frequencies of transitions $\ket{4}$-$\ket{5}$ and $\ket{4}$-$\ket{6}$, or $\frac{28.91+29.24}{2}=29.08$~GHz.  The lines at $\Delta_c/2\pi=\pm20$~MHz are the AT splitting caused by the RF1 field at a 1~mW power level. One of the interesting features here is the curving of the EIT lines occurring between $\Delta_{RF2}/2\pi=\pm200$~MHz.  At $\Delta_{RF2}/2\pi=-162.45$~MHz (or $f_{RF2}$=28.92~GHz, the $51P_{3/2}$-$51S_{1/2}$ transition), the EIT signal has a strong peak at $\Delta_c/2\pi=0$. If we detune further in the minus direction, the EIT line at $\Delta_{c}=0$ is red shifted and eventually disappears into the AT line at $\Delta_c/2\pi=-20$~MHz.  Similarly, at $\Delta_{RF2}/2\pi=162.45$~MHz (or $f_{RF2}$=29.24~GHz, the $51P_{3/2}$-$52S_{1/2}$ transition), the EIT signal has a strong peak at $\Delta_c=0$.  If we detune further in the plus direction, the EIT line at $\Delta_{c}=0$ is again red shifted and eventually disappears into the AT line at $\Delta_c/2\pi=-20$~MHz.
Now if we start at $\Delta_{RF2}/2\pi=-162.45$~MHz and increase $f_{RF2}$ to 29.24~GHz, we see that the EIT line is blue shifted and approaches the AT line at $\Delta_c/2\pi=20$~MHz. At $\Delta_{RF2}=0$, the EIT line turns-around and moves back toward $\Delta_c=0$ for increasing $\Delta_{RF2}$. The shape of the curve (or rate of this increase and decrease) is a function of the RF2 power level.

The point that the $m_J=\pm3/2$ magnetic sublevels in the $\ket{4}$ manifold are not coupled to $\ket{5}$ or to $\ket{6}$ is clearly illustrated in Fig.~\ref{rf2detuning} (columns 1 and 3), where see we straight lines beneath the bell-shaped curves. The straight lines are the AT lines for the $m_J=\pm3/2$ levels in the $\ket{4}$ manifold.

The bell-shaped behavior is further depicted in Fig.~\ref{quad}, where we show the location of the main peak EIT line ($\Delta_{c,peak}$) caused by the $\ket{4}$-$\ket{5}$ and $\ket{4}$-$\ket{5}$ transitions as a function of $\Delta_{RF2}$.
The zero crossing at $\Delta_{RF2}/2\pi=-162.45$~MHz corresponds to when RF2 is on-resonance with the $\ket{4}$-$\ket{5}$ transition and the zero crossing at $\Delta_{RF2}/2\pi=162.45$~MHz corresponds to when RF2 is on-resonance with the $\ket{4}$-$\ket{6}$ transition.
The curved behavior is due to the fact that when we detune RF2 we are either coupling to two transitions simultaneously or the coupling is dominated by one of the transitions ($\ket{4}$-$\ket{5}$ for $\Delta_{RF2}/2\pi<-162.45$~MHz or $\ket{4}$-$\ket{6}$ for $\Delta_{RF2}/2\pi>162.45$~MHz). Since the atomic dipole moments ($d$) are approximately equal for the $\ket{4}$-$\ket{5}$ and $\ket{4}$-$\ket{6}$ transitions, in the range between  $-162.45~{\rm MHz}<\Delta_{RF2}/2\pi<162.45~\rm{MHz}$ we are coupling to state $\ket{5}$ and $\ket{6}$ simultaneously with equal transition strengths. This is what causes the symmetric bell shaped behavior.   The separation distance between the two zero crossings is 324.8~MHz and is simply the difference in the on-resonant transitions frequencies of the two atomic transitions $\ket{4}$-$\ket{5}$ and $\ket{4}$-$\ket{6}$.

The error bars on Fig.~\ref{quad}
represent the standard deviation from ten sets of experiments,
indicating good repeatability of the measurement. The
errors and uncertainties associated with these measurements
are mainly due to laser power and laser frequency stability. While we have not done a detailed uncertainty analysis of these data sets, the errors and uncertainties of these types of measurements are related to the EIT/AT detection scheme
in general and are discussed in \cite{emc-conf}.

We see from the results in Fig.~\ref{quad}, that the value of $\Delta_{c,peak}$ (the coupling laser detuning frequency corresponding with a peak in the EIT transmittance) at $\Delta_{RF2}=0$ depends on the field strength of RF2. We also see that the slope of the curves at the zero crossing ($\Delta_{RF2}/2\pi=\pm162.45$~MHz) is also depends on the field strength of RF2. This is discussed in more detail in the next section.

%

\begin{figure*}
\centering
\scalebox{1.0}{\includegraphics{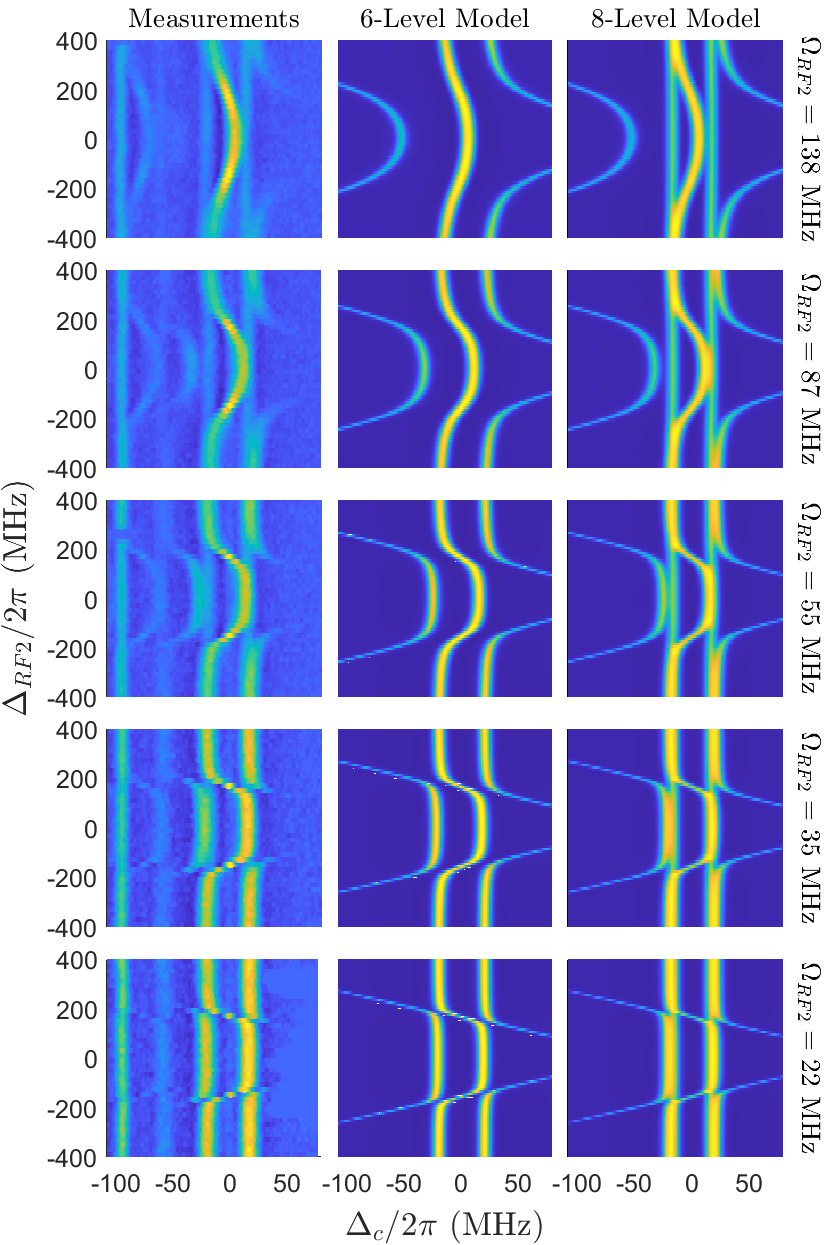}}
\caption{Experimental EIT signal for the six-level systems shown in Fig.~\ref{EIT}(b) for scanning the frequency of RF2 with $P_{RF1}=1$~mW [$\Omega_{RF1}=2\pi\times$(40~MHz)] for $P_{RF2}=10$~mW [$\Omega_{RF2}=2\pi\times$(138~MHz)],  $P_{RF2}=3.98$~mW [$\Omega_{RF2}=2\pi\times$(87~MHz)], $P_{RF2}=1.26$~mW [$\Omega_{RF2}=2\pi\times$(55~MHz)], $P_{RF2}=0.63$~mW [$\Omega_{RF2}=2\pi\times$(35~MHz)], and   $P_{RF2}=0.25$~mW [$\Omega_{RF2}=2\pi\times$(22~MHz). The left column are experimental results, the center column are simulations based on a 6-level model, and right column are simulations based on a 8-level model.}
\label{rf2detuning}
\end{figure*}

\begin{figure}
\vspace{5mm}
\centering
\scalebox{.35}{\includegraphics*{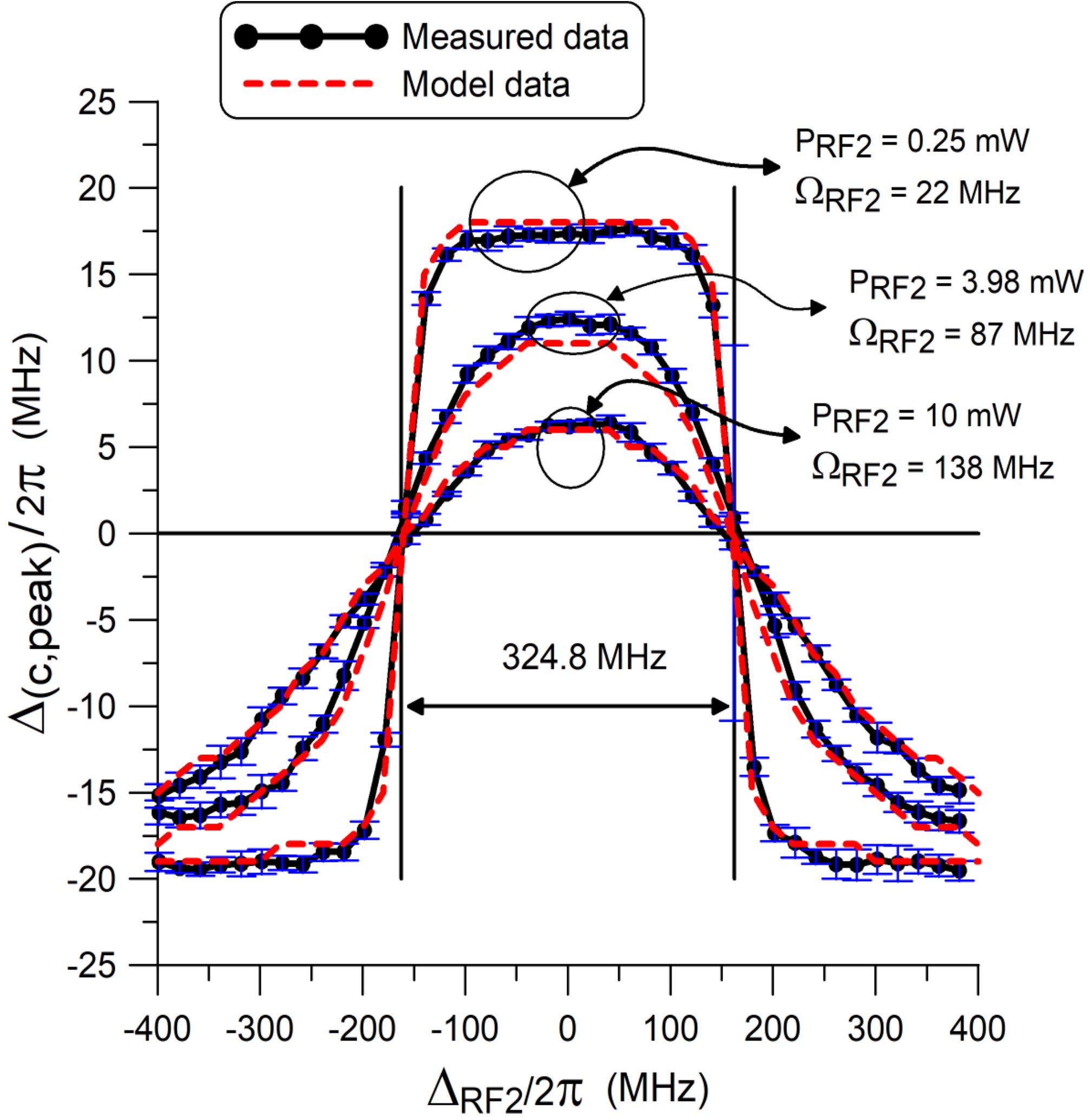}}\\
\caption{EIT peak location as function of RF2 detuning ($\Delta_{RF2}$) for $P_{RF1}=1$~mW [$\Omega_{RF1}=2\pi\times$(40~MHz)] with $P_{RF2}=10$~mW [$\Omega_{RF2}=2\pi\times$(138~MHz)], $P_{RF2}=3.98$~mW [$\Omega_{RF2}=2\pi\times$(87~MHz)], and $P_{RF2}=0.25$~mW [$\Omega_{RF2}=2\pi\times$(22~MHz)]. The
error bars represent the standard deviation of ten measurements.}
\label{quad}
\end{figure}

\section{Theoretical Model}

The most interesting features of this atomic system is the bell-shaped curves observed in the EIT line around $\Delta_c=0$~MHz for the case when RF2 is detuned. The model in this section will be used to simulate this bell-shaped behavior seen in the main EIT lines centered around $\Delta_c=0$~MHz corresponding to the six-level system shown in Fig.~\ref{EIT}(b).  In particular, we present a six-level model that will only represent the $m_J=\pm1/2$ magnetic sub-levels between the $\ket{4}$-$\ket{5}$ and $\ket{4}$-$\ket{6}$. This model will simulate the curved behavior in the EIT lines seen in Figs.~\ref{rf2detuning} and \ref{quad}. The straight lines under the curved line observed in the experimental data correspond to the $m_J=\pm3/2$ magnetic sub-levels. A higher-level model is required to duplicate all the features observed in the experimental data, including both the $m_J=\pm3/2$ and $m_J=\pm1/2$ magnetic sub-level. An eight-level model that captures most of these features is discussed in the appendix, which captures both the curved behavior of the EIT lines and the straight lines under the curved line seen in Fig~\ref{rf2detuning}.

In the six-level model, we use parameters for $^{85}{\rm Rb}$.
Similar results apply to other typical cases, such as,  $^{87}{\rm Rb}$ (which merely differs in partial vapor pressure) and $^{133}{\rm Cs}$.
We start by noting that the power of
the probe beam measured on the detector (the EIT signal, i.e., the probe transmission through the vapor cell)
is given by \cite{r15,berman}
\begin{equation}
P=P_0 \exp\left(-\frac{2\pi L \,\,{\rm Im}\left[\chi\right]}{\lambda_p}\right)=P_0 \exp\left(-\alpha L\right) \,\,\, ,
\label{intensity}
\end{equation}
where $P_0$ is the power of the probe beam at the input of the cell, $L$ is the length of the cell, $\lambda_p$ is the
wavelength of the probe laser,  $\chi$ is the susceptibility of the medium seen by the probe laser, and $\alpha=2\pi{\rm Im}\left[\chi\right]/\lambda_p$ is the
Beer's absorption coefficient for the probe laser.  The susceptibility
for the probe laser is related to the density matrix component
($\rho_{21}$) that is associated with the $\ket{1}$-$\ket{2}$ transition by the following \cite{berman}
\begin{equation}
\chi=\frac{2\,{\cal{N}}_0\wp_{12}}{E_p\epsilon_0} \rho_{21_D} =\frac{2\,{\cal{N}}_0}{\epsilon_0\hbar}\frac{(d\, e\, a_0)^2}{\Omega_p} \rho_{21_D}\,\,\, ,
\label{chi1}
\end{equation}
where $d=1.93$ is the normalized dipole moment for the probe laser, $\Omega_p$ is the Rabi frequency for the probe laser in units of rad/s. The subscript $D$ on $\rho_{12}$ presents a Doppler averaged value, ${\cal{N}}_0$ is the total density of atoms in the cell and is given by
\begin{equation}
{\cal{N}}_0=0.7217 \frac{p}{k_B T} \,\, ,
\label{nn}
\end{equation}
where $k_B$ is the Boltzmann constant, $T$ is temperature in Kelvin, and the pressure $p$ (in units of Pa) is given by \cite{stackrb}
\begin{equation}
p=10^{ 5.006+4.857-\frac{4215}{T}} \,\, .
\label{ppp}
\end{equation}
The factor 0.7217 in eq.~(\ref{nn}) reflects the natural abundance of $^{85}$Rb. In eq. (\ref{chi1}), $\wp_{12}$ is the dipole moment for the $\ket{1}$-$\ket{2}$ transition, $\epsilon_0$ is the vacuum permittivity, and $E_p$ is the amplitude of the probe laser E-field.

The density matrix component ($\rho_{21}$) is obtained from the master equation \cite{berman}
\begin{equation}
\dot{\rho}=\frac{\partial \rho}{\partial t}=-\frac{i}{\hbar}\left[H,\rho\right]+{\cal{L}} \,\,\, ,
\label{me}
\end{equation}
where $H$ is the Hamiltonian of the atomic system under consideration and ${\cal{L}}$ is the
Lindblad operator that accounts for the decay processes in the atom.

For the six-level system shown in Fig.~\ref{EIT}(b), the Hamiltonian can be expressed as:
\begin{equation}
\begin{footnotesize}
H=\frac{\hbar}{2}\left[\begin{array}{cccccc}
0 & \Omega_p & 0 & 0&0&0\\
\Omega_p & A & \Omega_c & 0&0&0\\
0 & \Omega_c & B & \Omega_{RF34}&0&0\\
0 & 0 & \Omega_{RF34} & C &\Omega_{RF45}&\Omega_{RF46}\\
0 & 0 &0& \Omega_{RF45} & D &0\\
0 & 0 &0& \Omega_{RF46} & 0 &E\\
\end{array}
\right]\,\, ,
\end{footnotesize}
\label{H4}
\end{equation}
where $RF_{34}$ is the RF source from $\ket{3}-\ket{4}$, $RF_{45}$ is the RF source from $\ket{4}-\ket{5}$, and $RF_{46}$ is the RF source from $\ket{4}-\ket{6}$. Also,
\begin{equation}
\begin{array}{rcl}
A&=&2\Delta_p \\
B&=&-2(\Delta_p+\Delta_c)\\
C&=&-2(\Delta_p+\Delta_c+\Delta_{RF34})\\
D&=&-2(\Delta_p+\Delta_c+\Delta_{RF34}-\Delta_{45})\\
E&=&-2(\Delta_p+\Delta_c+\Delta_{RF34}+\Delta_{46})\\
\end{array}
\end{equation}
where $\Delta_p$, $\Delta_c$, $\Delta_{RF34}$, and $\Delta_{4i}$ are the detunings of the probe laser, couple laser, the RF1 source, and the RF2 source, respectively; and $\Omega_p$, $\Omega_c$, $\Omega_{RF34}$, $\Omega_{RF45}$, and $\Omega_{RF46}$ are the Rabi frequencies associated with the probe laser,
coupling laser, and the RF sources. The additional minus sign in front of $\Delta_{45}$ is required because $\ket{5}$ is at a lower energy then $\ket{4}$, while $\ket{6}$ is at a higher energy than $\ket{4}$.  The detuning for each field is defined as
\begin{equation}
\Delta_{p,c,RFij}=\omega_{p,c,RFij}-\omega_{o_{p,c,RFij}} \,\,\, ,
\label{detuningeq}
\end{equation}
where $\omega_{o_{p,c,RFij}}$ are the on-resonance angular frequencies of transitions $\ket{1}$-$\ket{2}$, $\ket{2}$-$\ket{3}$, and $\ket{i}$-$\ket{j}$, respectively;
and $\omega_{p,c,RFij}$ are the angular frequencies of the probe laser, coupling laser, and the RF source, respectively.

In our experiments, $\Delta_p=0$, $\Delta_{RF34}=0$, $\Delta_c$ is scanned, and
\begin{equation}
\begin{array}{rcl}
\Delta_{45}&=&2\pi\cdot[f_{RF2}-28.92~{\rm GHz}]\\
\Delta_{46}&=&2\pi\cdot[f_{RF2}-29.24~{\rm GHz}]\\
\end{array}
\end{equation}
where $f_{RF2}$ is the frequency of the RF2 source.

For the six-level system, the ${\cal{L}}$ matrix is given in eq.~(\ref{L4}),
\begin{figure*}
\begin{equation}
\begin{footnotesize}
{\cal{L}}=\left[\begin{array}{cccccc}
\Gamma_2 \rho_{22} & -\gamma_{12}\rho_{12} & -\gamma_{13}\rho_{13} & -\gamma_{14}\rho_{14}& -\gamma_{15}\rho_{15}  & -\gamma_{16}\rho_{16} \\

-\gamma_{21}\rho_{21} & \Gamma_3 \rho_{33}-\Gamma_2 \rho_{22} & -\gamma_{23}\rho_{23} & -\gamma_{24}\rho_{24}& -\gamma_{25}\rho_{25}  & -\gamma_{26}\rho_{26}\\

-\gamma_{31}\rho_{31} & -\gamma_{32}\rho_{32} & \Gamma_4 \rho_{44}-\Gamma_3 \rho_{33} & -\gamma_{34}\rho_{34}& -\gamma_{35}\rho_{35}  & -\gamma_{36}\rho_{36}\\

-\gamma_{41}\rho_{41} & -\gamma_{42}\rho_{42} & -\gamma_{43}\rho_{43} &  \Gamma_5 \rho_{55}+\Gamma_6 \rho_{66}-\Gamma_4 \rho_{44}& -\gamma_{45}\rho_{45}  & -\gamma_{46}\rho_{46}\\

-\gamma_{51}\rho_{51} & -\gamma_{52}\rho_{52} & -\gamma_{53}\rho_{53} & -\gamma_{45}\rho_{45}  &  -\Gamma_5 \rho_{55}& -\gamma_{56}\rho_{56}\\

-\gamma_{61}\rho_{61} & -\gamma_{62}\rho_{62} & -\gamma_{63}\rho_{63} & -\gamma_{65}\rho_{65} & -\gamma_{56}\rho_{56} &  -\Gamma_6 \rho_{66}\\

\end{array}
\right] \,\,\, ,
\end{footnotesize}
\label{L4}
\end{equation}
\end{figure*}
where $\gamma_{ij}=(\Gamma_i+\Gamma_j)/2$ and $\Gamma_{i, j}$ are the transition decay rates. Since the purpose of the present study is to explore the intrinsic limitations of Rydberg-EIT field sensing in vapor cells, no collision terms or dephasing terms
are added. While Rydberg-atom collisions, Penning ionization, and ion electric fields can, in principle, cause dephasing, such effects can, for instance, be alleviated by reducing
the beam intensities, lowering the vapor pressure, or limiting the atom-field interaction time. In this analysis we set,
$\Gamma_1=0$, $\Gamma_2=2\pi\times$(6~{\rm MHz}),
$\Gamma_3=2\pi\times$(3~{\rm kHz}), $\Gamma_{4,5,6}=2\pi\times$(2~{\rm kHz}).
Note, $\Gamma_{2}$ is for the D2 line in $^{85}$Rb \cite{stackrb}, and $\Gamma_{3}$, $\Gamma_{4,5,6}$, are typical Rydberg decay rates.

We numerically solve these equations to find the steady-state solution for $\rho_{21}$ for various values of $\Omega_c$, $\Omega_p$, and $\Omega_{RFij}$.
This is done by forming a matrix with the system of equations for $\dot{\rho}_{ij}=0$. The null-space of this resulting system matrix is the
steady-state solution.  The steady-state solution for $\rho_{21}$ is Doppler averaged in the usual way
\begin{equation}
\rho_{21_D}=\frac{1}{\sqrt{\pi}\,\, u}\int_{-3u}^{3u}\rho_{21}\left(\Delta'_p,\Delta'_c\right)\,\,e^{\frac{-v^2}{u^2}}\,\,dv\,\,\, ,
\label{doppler}
\end{equation}
where $u=\sqrt{2k_B T/m}$ and $m$ is the mass of the atom. We use the case where the probe and coupling laser are counter-propagating.
Thus, the frequency seen by the atom moving toward the probe beam is upshifted by $2\pi v/\lambda_p$ (where $v$ is the velocity of the atoms),
while the coupling beam is downshifted by $2\pi v/\lambda_c$.  Thus, the probe and coupling beam detuning is modified by the following
\begin{equation}
\Delta'_p=\Delta_p-\frac{2\pi}{\lambda_p}v \,\,\,{\rm and}\,\,\,
\Delta'_c=\Delta_c+\frac{2\pi}{\lambda_c}v \,\,\, .
\label{doppler2}
\end{equation}

Fig.~\ref{rf2detuning}(the center column) shows the EIT spectra obtained from the six-level model for the same parameters used for the experimental data (the left colum of Fig.~\ref{rf2detuning}). When comparing the results in these two columns (the left and center columns) we see very similar behavior is the spectra. In that, the six-level model captures the curved behavior as that observed in the experimental data. However, the straight lines under the curved lines in Fig.~\ref{rf2detuning} are due to the $m_J=\pm3/2$ magnetic sub-levels, which is not captured in this six-level model.
A model for capturing the EIT spectra for the $m_J=\pm3/2$ magnetic sublevels is discussed in the appendix.


\begin{figure}
\vspace{5mm}
\centering
\scalebox{.30}{\includegraphics*{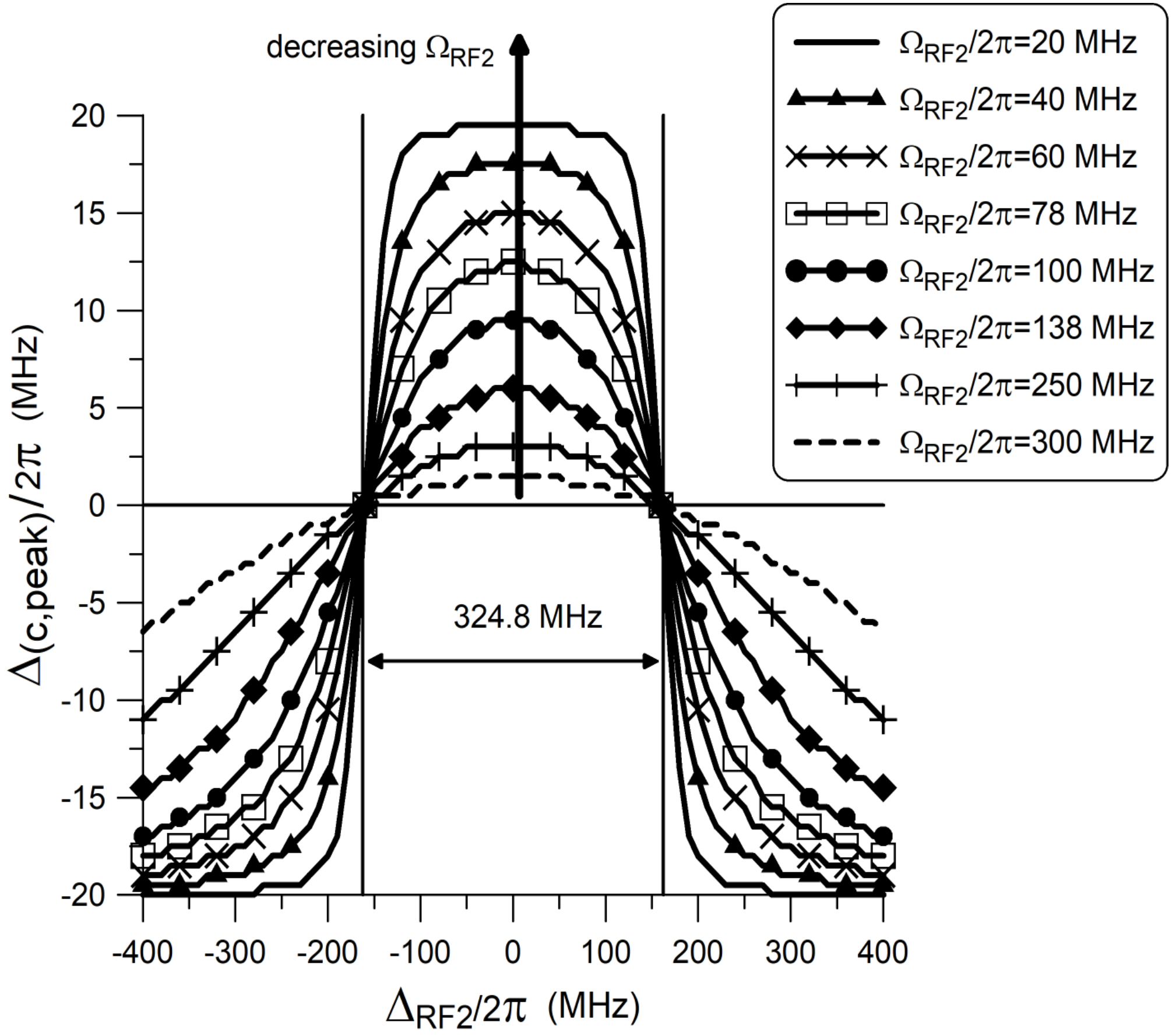}}\\
{\hspace{-1mm}\tiny{(a) $\Omega_{RF1}=2\pi\times$(40~MHz)}}\\
\scalebox{.30}{\includegraphics*{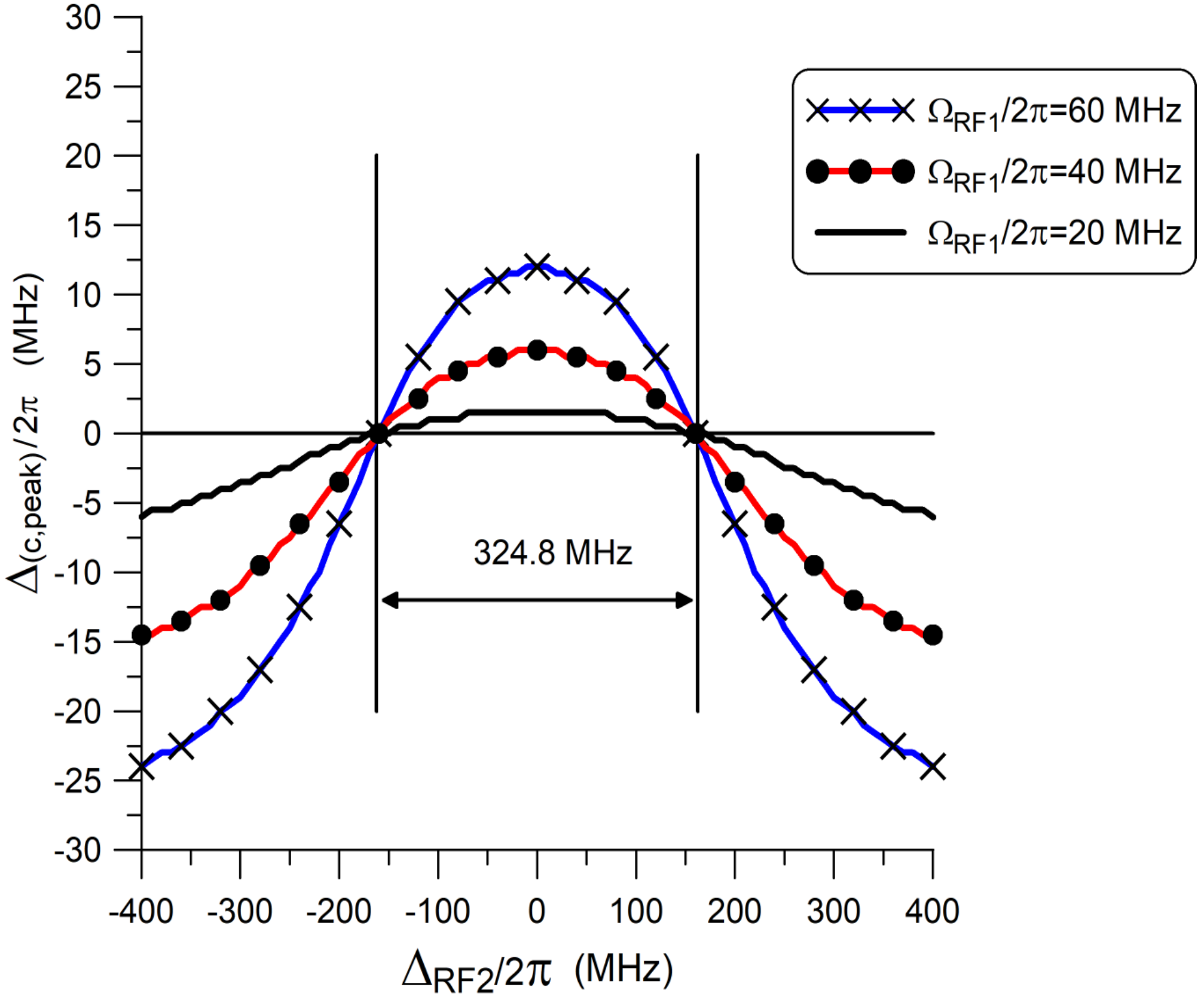}}\\
{\hspace{-1mm}\tiny{(b) $\Omega_{RF2}=2\pi\times$(138~MHz)}}\\
\scalebox{.30}{\includegraphics*{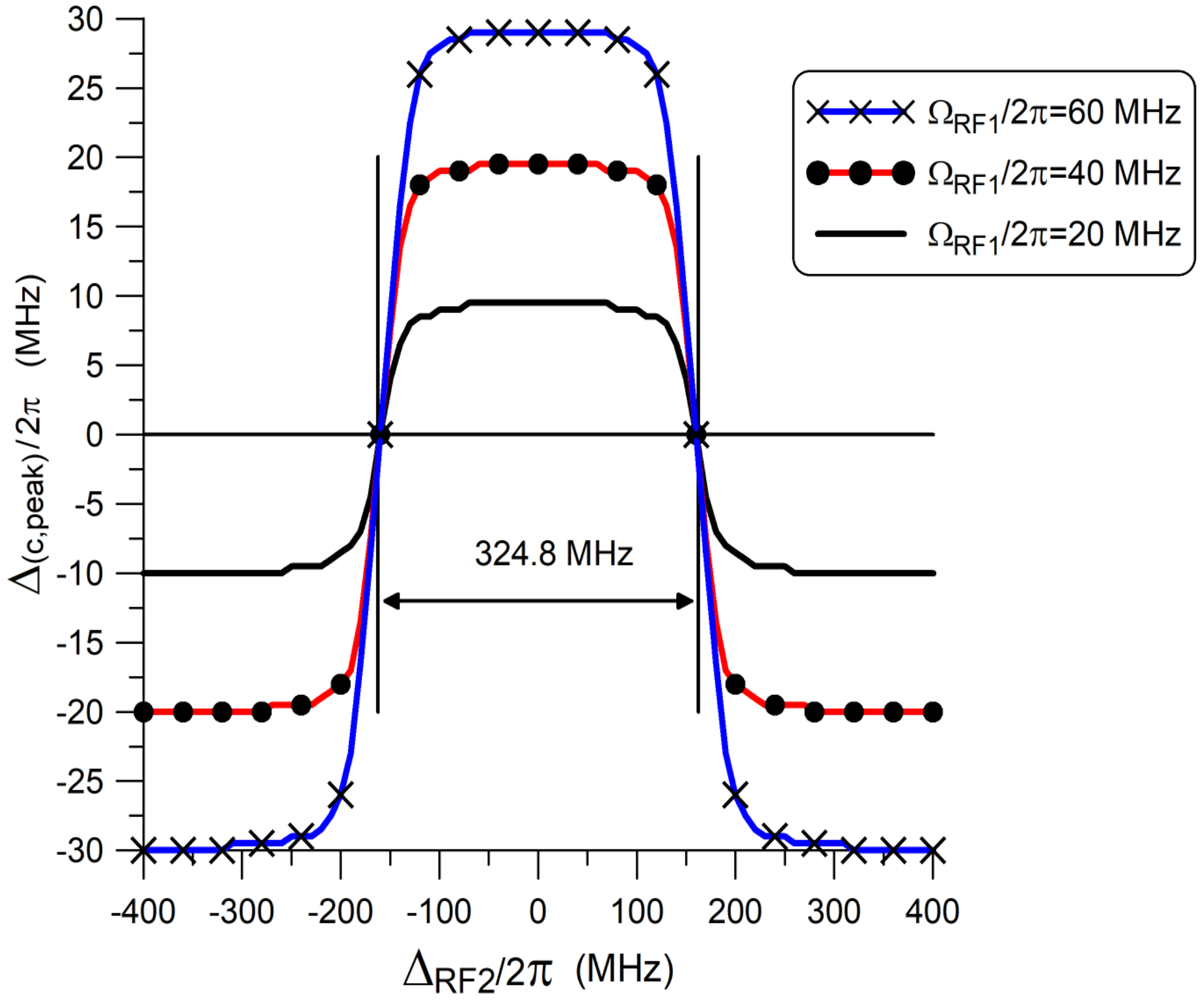}}\\
{\hspace{-1mm}\tiny{(c) $\Omega_{RF2}=2\pi\times$(20~MHz)}}\\
\caption{Modeled results for the EIT peak location as function of RF2 detuning ($\Delta_{RF2}$) for various values of $\Omega_{RF2}$ (or $P_{RF2}$) and $\Omega_{RF1}$ (or $P_{RF1}$): (a) $\Omega_{RF1}=2\pi\times$(40~MHz), (b) $\Omega_{RF2}=2\pi\times$(138~MHz),
(c) $\Omega_{RF2}=2\pi\times$(20~MHz).}
\label{model2}
\end{figure}

Using the six-level model, we generated a family of plots for the location of the peak EIT line caused by the $\ket{4}$-$\ket{5}$ and $\ket{4}$-$\ket{6}$ transition as a function of $\Delta_{RF2}$ (similar to the data shown in Fig.~\ref{quad}). This family of plots is shown in Fig.~\ref{model2}. Fig.~\ref{model2}(a) is for $\Omega_{RF1}/2\pi=40$~MHz for various values of $\Omega_{RF2}$, Fig.~\ref{model2}(b) is for three values of $\Omega_{RF1}$ for $\Omega_{RF2}/2\pi=136$~MHz, and Fig.~\ref{model2}(c) is for three values of $\Omega_{RF1}$ for $\Omega_{RF2}/2\pi=22$~MHz.  We see very similar behavior as that shown in Fig.~\ref{quad}.  In fact, in Fig.~\ref{quad}, we made a comparison of the experimental data to the model.  This figure shows very good agreement between the experiments and model. This illustrates that the six-level model can accurately capture the bell-shaped behavior of the EIT lines.

There are two interesting things to observe in the data shown in Figs.~\ref{quad} and \ref{model2}.  First, when $\Delta_{RF2}$ is detuned, $\Delta_{c,peak}$ reaches a maximum at $\Delta_{RF2}=0$. Furthermore, the maximum is a function of $\Omega_{RF2}$ (the amplitude of the applied RF2 source).  Secondly, all the plots cross at $\Delta_{c,peak}=0$ at two points ($\Delta_{RF2}/2\pi=\pm162.45$~MHz). At the crossing we see that the slope of the curves is dependant on $\Omega_{RF2}$ (the amplitude of the applied RF2 source).
These two features suggest that these behaviors can be used to infer the field strength of RF2.  Fig.~\ref{peak1}(a) shows the value of $\Delta_c$ that corresponds with the top of the bell-shape versus $\Omega_{RF2}$ for three different $\Omega_{RF1}$. Fig.~\ref{peak1}(b) shows the slope at the zero crossing ($\Delta_{RF2}/2\pi=162.45$~MHz) versus $\Omega_{RF2}$ for different $\Omega_{RF1}$. The results in these figures indicate that by determining either the peak value or the slope could be a method for inferring amplitude of the field strength of RF2.

\begin{figure}
\vspace{5mm}
\centering
\scalebox{.30}{\includegraphics*{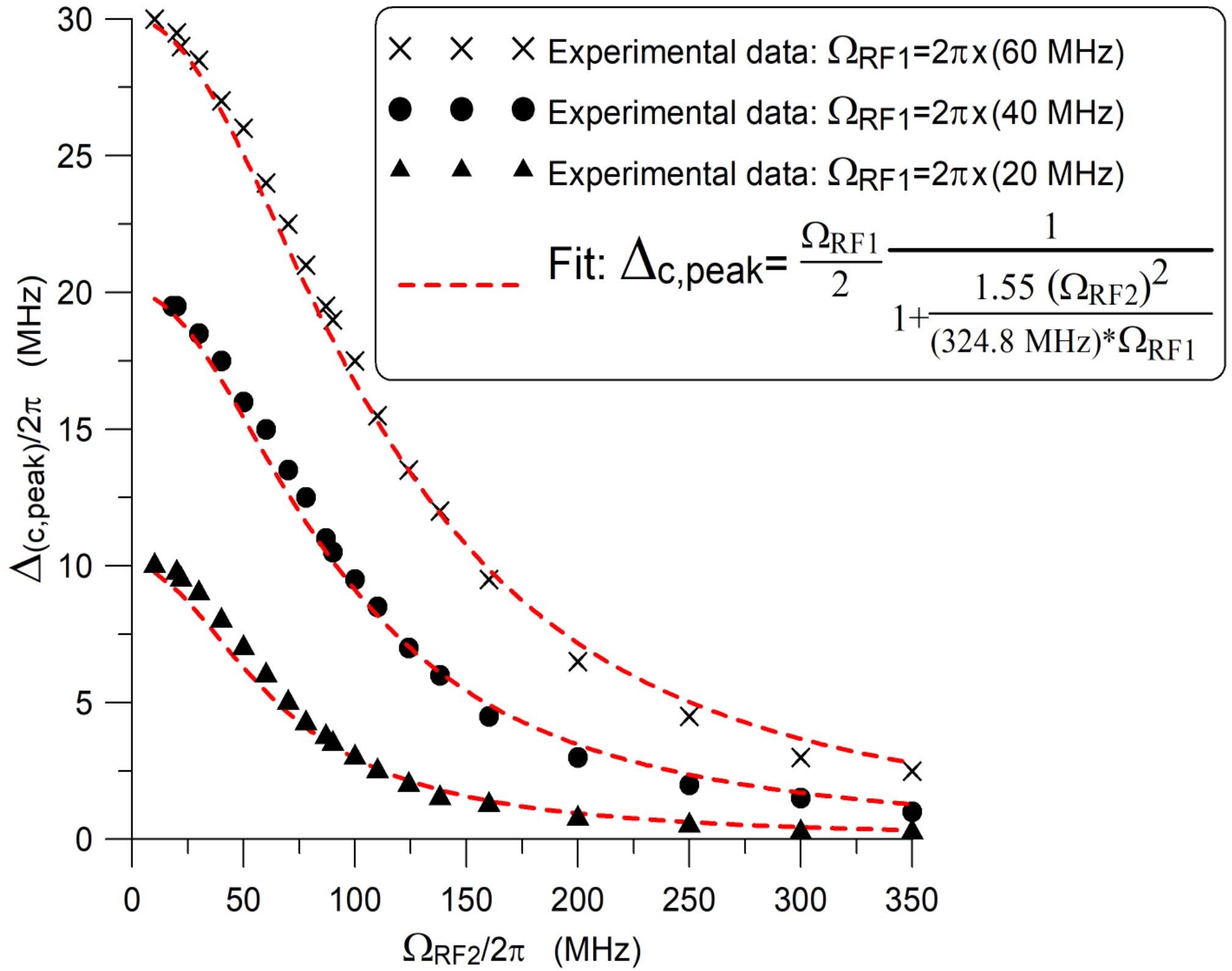}}\\
{\hspace{-1mm}\tiny{(a) Peak location}}\\
\scalebox{.38}{\includegraphics*{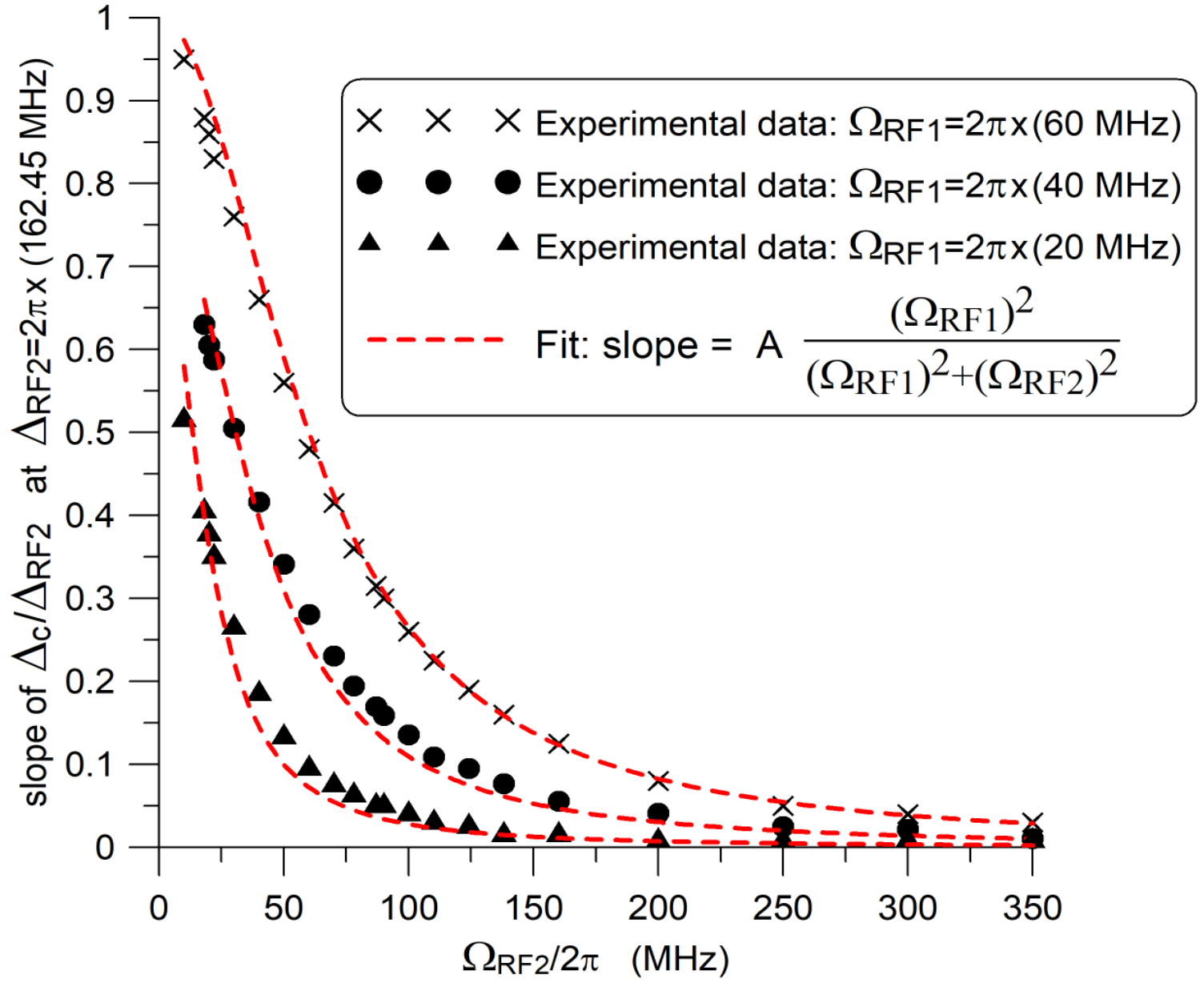}}\\
{\hspace{-1mm}\tiny{(a) Slope}}\\
\caption{Modeled results for: (a) the value of $\Delta_c$ where the EIT peak location as function of $\Omega_{RF2}$ for various values of $\Omega_{RF1}$ (or $P_{RF1}$), (b) the slope at $\Delta_{RF2}/2\pi=162.45$~MHz as function of $\Omega_{RF2}$ for various values of $\Omega_{RF1}$ (or $P_{RF1}$).}
\label{peak1}
\end{figure}

The data in these two figures can be fit to empirical formulas as given below:
\begin{equation}
\Delta_{c, peak}=\frac{\Omega_{RF1}}{2}\frac{1}{1+1.55\frac{\left(\Omega_{RF2}\right)^2}{\Omega_{RF1} \,\,\,2\pi\times(324.8~{\rm MHz})}}\\
\label{fits1}
\end{equation}
\begin{equation}
Slope=A\frac{\left(\Omega_{RF1}\right)^2}{\left(\Omega_{RF1}\right)^2+\left(\Omega_{RF2}\right)^2}\\
\label{fit2s}
\end{equation}
where $A$ is a fit parameter: $A=1.0$ for $\Omega_{RF2}/2\pi=60$~MHz,
$A=0.8$ for $\Omega_{RF2}/2\pi=40$~MHz, and $A=0.7$ for $\Omega_{RF2}/2\pi=20$~MHz. These two fits are also show in Fig.~\ref{peak1}.
These expressions can be used to infer the E-field strength of RF2. Future work will explore these expressions for a range of other parameters and investigate potential applications.

Furthermore, inferring the strength ratio of RF1/RF2 could also be an application of the power sweeps in Fig.~\ref{rf2power}. We nearly saturate our AT split with $P_{RF1}$=10~mW, but the rate that the RF2 EIT center peak appears to change with RF1 power. One could infer a saturated RF power by applying that RF2 field. This will be a topic of future work.

\section{Discussion}

In this paper, we investigated a six-level EIT scheme in Rydberg atoms, which consisted of two lasers (a probe laser and a coupling laser), and two RF sources. We chose states such that one RF source could simultaneously couple to two different Rydberg states. This resulted in a more complicated EIT spectrum with interesting features (when compared to the standard four-level scheme useD in electric field sensing).  We developed models to analyze this atomic system, and showed that the models accurately simulate the experimental data. In this work, we only explored coupling to two similar states of the same angular momentum. Future work will explore multi-level EIT spectra with different angular momentum couplings.

As Rydberg atom EIT is currently being investigated for RF field sensing, it is important to understand how the spectra change when multiple RF fields in different frequency bands are present. This more complicated atomic system may also be used as an alternative way to measure the field strength of the second applied RF field. We discussed two methods for estimating the E-field strength of the second applied RF field, by presenting empirical formulas to relate the peak EIT peak location and slope to the field strength of the second RF source. In future work, we will explore using these two sensing techniques, as well as explore other aspects of this multi-level EIT system.  The multi-level schemes can allow for additional sensing capabilities, as well as to evaluate atom-based receivers in the presence of multiple RF signals.

\section{Acknowledgement}
The authors thank Prof. G. Raithel with the University of Michigan and Rydberg Technologies LLC, Ann Arbor, MI, for his useful and insightful technology discussions.


\appendix
\section{Eight-Level Modal for both $m_J=\pm1/2$ and $m_J=\pm3/2$
Magnetic Sub-Level}

The following will model the EIT lines corresponding to the eight-level system shown in Fig.~\ref{EIT8}. This model extends the six-level model presented in section IV, in that, $\ket{1}$ through $\ket{6}$ are the same as used in the six-level system, and $\ket{7}$ corresponds to the $m_J=\pm3/2$ magnetic sub-level transition from $\ket{2}$, and $\ket{8}$ corresponds to the $m_J=\pm3/2$ magnetic sub-level transition from $\ket{7}$.  The $\ket{2}$-$\ket{7}$-$\ket{8}$ transition corresponds to the straight lines beneath the bell-shaped curve observed in the experimental data shown in left column of Fig.~\ref{rf2detuning}.    Note that because transitions $\ket{4}$-$\ket{5}$ and $\ket{4}$-$\ket{6}$ are P-S transitions, the magnetic sub-levels $m_J=\pm3/2$ are not allowed. Therefore in this model, the transition from $\ket{2}$-to-$\ket{7}$ then to $\ket{7}$-to-$\ket{8}$ accounts for the $m_J=\pm3/2$ magnetic sub-levels and the transition from $\ket{3}$-to-$\ket{4}$ accounts for the $m_J=\pm1/2$ magnetic sub-levels; where the $m_J=\pm1/2$ sub-levels are couple to $\ket{5}$ and $\ket{6}$.

\begin{figure}
\centering
\scalebox{.4}{\includegraphics*{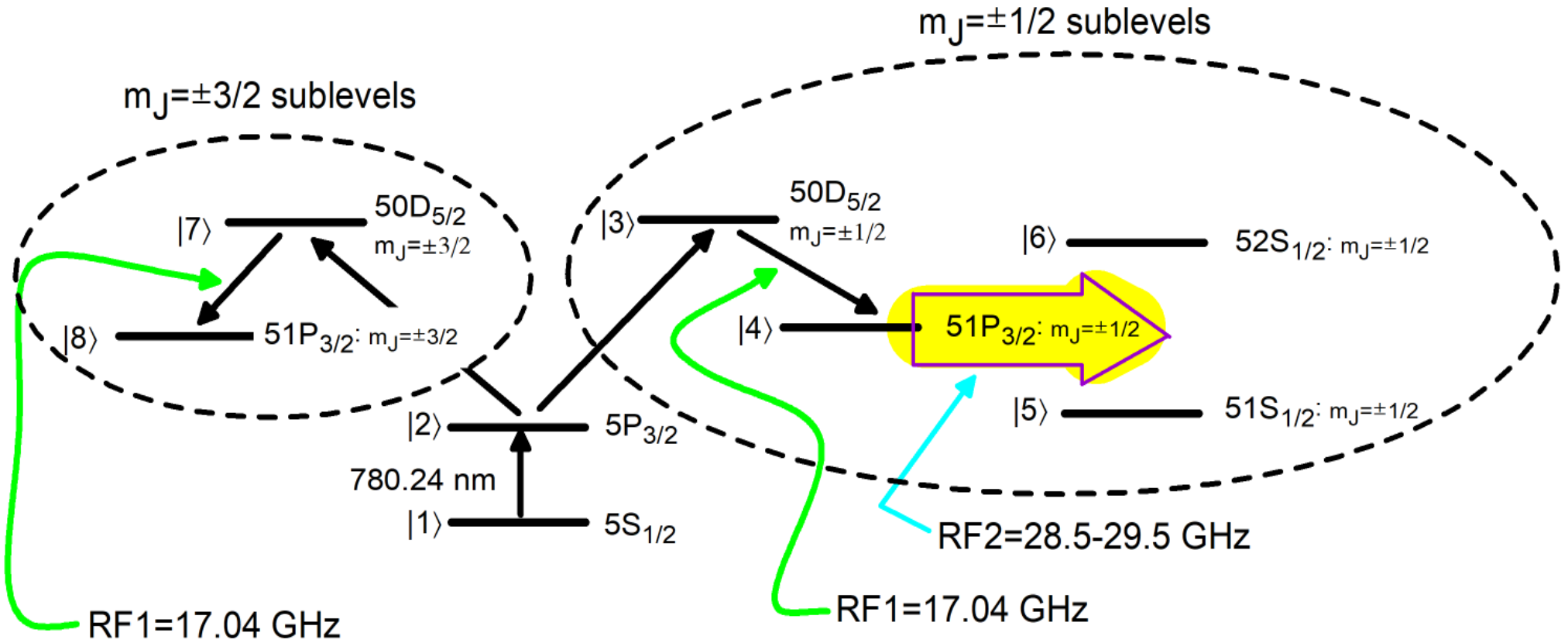}}\\
\caption{Eight -level system for modeling both $m_J=\pm1/2$ and $m_J=\pm3/2$ magnetic sub-levels.}
\label{EIT8}
\end{figure}

For the eight-level system, the Hamiltonian can be expressed as in eq.~(\ref{H48}),
\begin{figure*}
\begin{equation}
\begin{footnotesize}
H=\frac{\hbar}{2}\left[\begin{array}{cccccccc}
0 & \Omega_p & 0 & 0&0&0&0&0\\
\Omega_p & A & \Omega_c & 0&0&0&\Omega_{RF27}&0\\
0 & \Omega_c & B & \Omega_{RF34}&0&0&0&0\\
0 & 0 & \Omega_{RF34} & C &\Omega_{RF45}&\Omega_{RF46}&0&0\\
0 & 0 &0& \Omega_{RF45} & D &0&0&0\\
0 & 0 &0& \Omega_{RF46} & 0 &E&0&0\\
0 & \Omega_{RF27} &0& 0 & 0 &0&F&\Omega_{RF78}\\
0 & 0 &0& 0 & 0 &0&\Omega_{RF78}&G\\
\end{array}
\right]\,\, ,
\end{footnotesize}
\label{H48}
\end{equation}
\end{figure*}
where $RF_{34}$ is the RF source from $\ket{3}-\ket{4}$, $RF_{45}$ is the RF source from $\ket{4}-\ket{5}$, and $RF_{46}$ is the RF source from $\ket{4}-\ket{6}$.
Note that $RF_{27}$ is the RF source from $\ket{2}-\ket{7}$, and is the same as the source from $\ket{2}-\ket{3}$ (the coupling laser), but $\Omega_{RF27}$ is scaled by the ratio of the angular-part of the dipole moment (see below).  $RF_{78}$ is the RF source from $\ket{7}-\ket{8}$, and is the same as the source from $\ket{3}-\ket{4}$, but $\Omega_{RF78}$ is scaled by the ratio of the angular-part of the dipole moment (see below).
Also,
\begin{equation}
\begin{array}{rcl}
A&=&-2\Delta_p \\
B&=&-2(\Delta_p+\Delta_c)\\
C&=&-2(\Delta_p+\Delta_c+\Delta_{RF34})\\
D&=&-2(\Delta_p+\Delta_c+\Delta_{RF34}-\Delta_{45})\\
E&=&-2(\Delta_p+\Delta_c+\Delta_{RF34}+\Delta_{46})\\
F&=&-2(\Delta_p+\Delta_c)\\
G&=&-2(\Delta_p+\Delta_c+\Delta_{RF78})\\
\end{array}
\end{equation}
where $\Delta_p$, $\Delta_c$, and $\Delta_{RF34}$ are the detunings of the probe laser, coupling laser, and the RF1 source,
respectively; and $\Omega_p$, $\Omega_c$, $\Omega_{RF34}$, $\Omega_{RF45}$, and $\Omega_{RF46}$ are the Rabi frequencies associated with the probe laser,
coupling laser, and the RF sources. The detuning for each field is defined as in eq.~(\ref{detuningeq}) in the main text.

In our experiments, $\Delta_p=0$, $\Delta_c$ will be scanned, $\Delta_{RF34}=0$, $\Delta_{RF78}=0$, and
\begin{equation}
\begin{array}{rcl}
\Delta_{45}&=&2\pi\cdot[f_{RF2}-28.92~GHz]\\
\Delta_{46}&=&2\pi\cdot[f_{RF2}-29.24~GHz]\\
\end{array}
\end{equation}

For the eight-level system, the ${\cal{L}}$ matrix is given in eq.~(\ref{H4L8}),
\begin{figure*}
\begin{equation}
\begin{footnotesize}
{\cal{L}}=\left[\begin{array}{cccccccc}
\Gamma_2 \rho_{22} & -\gamma_{12}\rho_{12} & -\gamma_{13}\rho_{13} & -\gamma_{14}\rho_{14}& -\gamma_{15}\rho_{15}  & -\gamma_{16}\rho_{16} & -\gamma_{17}\rho_{17}& -\gamma_{18}\rho_{18} \\

-\gamma_{21}\rho_{21} & \Gamma_3 \rho_{33}+\Gamma_7 \rho_{77}-\Gamma_2 \rho_{22} & -\gamma_{23}\rho_{23} & -\gamma_{24}\rho_{24}& -\gamma_{25}\rho_{25}  & -\gamma_{26}\rho_{26}& -\gamma_{27}\rho_{27}& -\gamma_{28}\rho_{28}\\

-\gamma_{31}\rho_{31} & -\gamma_{32}\rho_{32} & \Gamma_4 \rho_{44}-\Gamma_3 \rho_{33} & -\gamma_{34}\rho_{34}& -\gamma_{35}\rho_{35}  & -\gamma_{36}\rho_{36}& -\gamma_{37}\rho_{37}& -\gamma_{38}\rho_{38}\\

-\gamma_{41}\rho_{41} & -\gamma_{42}\rho_{42} & -\gamma_{43}\rho_{43} &  \Gamma_5 \rho_{55}+\Gamma_6 \rho_{66}-\Gamma_4 \rho_{44}& -\gamma_{45}\rho_{45}  & -\gamma_{46}\rho_{46}& -\gamma_{47}\rho_{47} & -\gamma_{48}\rho_{48}\\

-\gamma_{51}\rho_{51} & -\gamma_{52}\rho_{52} & -\gamma_{53}\rho_{53} & -\gamma_{45}\rho_{45}  &  -\Gamma_5 \rho_{55}& -\gamma_{56}\rho_{56}& -\gamma_{57}\rho_{57} & -\gamma_{58}\rho_{58}\\

-\gamma_{61}\rho_{61} & -\gamma_{62}\rho_{62} & -\gamma_{63}\rho_{63} & -\gamma_{65}\rho_{65} & -\gamma_{56}\rho_{56} &  -\Gamma_6 \rho_{66}& -\gamma_{67}\rho_{67} & -\gamma_{68}\rho_{68}\\

-\gamma_{71}\rho_{71} & -\gamma_{72}\rho_{72} & -\gamma_{73}\rho_{73} & -\gamma_{74}\rho_{74} & -\gamma_{75}\rho_{75} & -\gamma_{76}\rho_{76}&  \Gamma_8 \rho_{88}-\Gamma_7 \rho_{77}& -\gamma_{78}\rho_{78}\\

-\gamma_{81}\rho_{81} & -\gamma_{82}\rho_{82} & -\gamma_{83}\rho_{83} & -\gamma_{84}\rho_{84} & -\gamma_{85}\rho_{85} & -\gamma_{86}\rho_{86}&  -\gamma_{87}\rho_{87} & -\Gamma_8 \rho_{88}\\

\end{array}
\right] \,\,\, ,
\end{footnotesize}
\label{H4L8}
\end{equation}
\end{figure*}
where $\gamma_{ij}=(\Gamma_i+\Gamma_j)/2$ and $\Gamma_{i, j}$ are the transition decay rates.  In this analysis we set $\Gamma_1=0$, $\Gamma_2=2\pi\times$(6~MHz),
$\Gamma_{3,7}=2\pi\times$(3~kHz), and $\Gamma_{(4,5,6,8)}=2\pi\times$(2~kHz).
Note, $\Gamma_{2}$ is for the D2 line in $^{85}$Rb \cite{stackrb}, and $\Gamma_{3,7}$, $\Gamma_{4,5,6,8}$, are typical Rydberg decay rates.

In this analysis,  we set $\Omega_{RF27}$ to $82~\%$ of $\Omega_{c}$ (or $\Omega_{RF27}=0.82\cdot\Omega_{c}$)
and $\Omega_{RF78}$ to $82~\%$ of $\Omega_{RF34}$ (or $\Omega_{RF78}=0.82\cdot\Omega_{RF34}$) . This factor 0.82 is the ratio of the angular part of the diploe moment for $m_J=1/2$ and $m_J=3/2$, or 0.4/.48989=0.82.

The right column of Fig.~\ref{rf2detuning} shows the results for the eight-level systems. Upon comparing to the results in the left column of Fig.~\ref{rf2detuning} we see that this eight-level model correlates very well to the experimental data, in that it captures the EIT lines associated with both the $m_J=\pm1/2$  and $m_J=\pm3/2$  magnetic sub-levels. When comparing the eight-level model to the six-level model (center column of Fig.~\ref{rf2detuning}),  the eight-level captures the $m_J=\pm3/2$ magnetic sublevels (the straight under the curved lines), where the six-level does not.  In this paper, we are more interested in the EIT structure associated with the $m_J=\pm1/2$ transitions, as such the six-level is sufficient in our analysis and use used in the main part of the paper. But it is worth noting that the eight-level model captures the magnetic sublevel behaviors observed in the experiments at $\Delta_c/2\pi=\pm20$~MHz. Other $m_J=\pm1/2$ and $m_{J}=\pm3/2$ magnetic sub-level features are observed in the experimental data around $\Delta_c/2\pi=\pm60$~MHz and $\Delta_c/2\pi=\pm100$~MHz that are not captures by the eight-level model. These other features are associated with the fine and hyper-fine structures, and can be included if desired by simply adding more complexity by way of more levels to the simulations.


\end{document}